# Understanding the Information Needs and Practices of Human Supporters of an Online Mental Health Intervention to Inform Machine Learning Applications

Anja Thieme
Microsoft Research
(September 2020)

# 1. Introduction

This research forms part of Project Talia (https://www.microsoft.com/en-us/research/project/project-talia/), a collaborative effort that aims to identify what might constitute effective and appropriately designed machine learning (ML) applications for mental health. More specifically, our work focuses on existing digital mental health services that deliver internet-based Cognitive Behavioral Therapy (iCBT) treatment to people who suffer from symptoms of depression and anxiety.

CBT is one of the most widely applied and extensively empirically tested psychotherapy programs in Western Healthcare [4]. Due to its highly structured format – that teaches the person to attend to the relationships between their thoughts, feelings and behaviors – it can be readily supported by digital technology [5, 6, 19]. Extensive research has further evidenced the clinical effectiveness of iCBT interventions for attaining sustainable improvements in people's mental health that are comparable to face-to-face therapy [1, 2, 21, 22]. Despite these advantages, digital behavioral interventions such as iCBT can suffer from high rates of attrition [8, 10, 11] meaning that users may not sustain engagement with treatment and are at risk of dropping-out from therapy before they may get the desired benefits. Here, extensive research has shown how the involvement of a human supporter or coach, who provides guidance and assistance to the person through online messages, improves user engagement in therapy and leads to more effective mental health outcomes than unsupported interventions [5,7, 9, 17, 22].

Seeking to maximize the effects and outcomes of supporter involvement, our research investigates how ML-enabled insights could support the work practices of iCBT supporters. More specifically, we explore how providing them with key data insights can: (i) enable more efficient clinical decision making; (ii) promote better treatment personalisation; and (iii) assist in more effective supporter communications and training.

We explore these opportunities in close collaboration with researchers and developers at SilverCloud Health (www.silvercloudhealth.com), which is an established iCBT platform for the treatment of depression, anxiety, and functional impairments. The platform offers a wide range of self-guided iCBT treatment programs whose clinical effectiveness has been evidenced through extensive clinical research, including randomized clinical trials (*e.g.*, [16]). Each program contains a set of core psycho-educational and psycho-therapeutic modules that are delivered using textual, video and audio contents as well as interactive activities, tools, quizzes and personal stories [5]. While it is recommended to complete one module each week, clients can work through the program content at their own pace and time.

To promote continued use and engagement, clients receive support from a trained supporter in the form of weekly reviews throughout their treatment journey. Most often these supporters are graduate psychologists who received further training in low-intensity interventions that are CBT based, including iCBT. As a qualified occupation, they take on the role of a Psychological Wellbeing Practitioner (PWP) [15], who undertakes risk assessment and offers evidence-based interventions to clients with mild-to-moderate symptoms of depression and anxiety within the NHS England IAPT initiative (Improving Access to Psychological Therapies) [14]. To assist clients who are using SilverCloud programs, their support typically involves the writing of personalized feedback messages to the client on their work, which usually takes 10-15 minutes to complete.

To better understand the existing work practices and information needs of iCBT supporters to inform the development of useful and implementable ML applications for this context, we first conducted semi-structured with 15 Psychological Wellbeing Practitioners. Seeking to better understand how these supporters currently gain an understanding of their clients' behavior and mental health outcomes from the data that is available to them via the online intervention (see Figure 3 for an example); and how they use this information as part of their support practices, our research makes two contributions:

(1) We derived insights about key information challenges that iCBT supporters encounter and the strategies that they employ for providing effective, personalized assistance as part of their current work practices.
(2) Responding to these learnings we started to scope out potential opportunities for ML to help derive and support identified information needs.

## 2. Background: PWP Support Process with SilverCloud

In the UK, the National Institute for Health and Care Excellence[1] (NICE) provides national guidance and advice to improve health and social care. For adult primary care of common mental health problems such as anxiety and depression, NICE guidelines support the use of CBT and introduced a stepped care model to improve treatment efficiency [18]. This care model ensures that the most effective, yet least resource intensive treatment is delivered first, and care is only stepped-up to more intensive face-to-face treatments if required [20]. In 2008, the NHS introduced the Improving Access to Psychological Therapies (IAPT) programme [14] specifically to help people overcome depression and anxiety and to better manage their mental. Figure 1 shows the stepped care model that IAPT services use to make clinical decisions on the most appropriate treatment for a specific client [12]. Psychological Wellbeing Practitioners (PWPs) work a step 2 within IAPT services, which offers access to SilverCloud Health programmes for the treatment of clients who typically suffer from mild-to-moderate symptoms of depression, anxiety or stress.

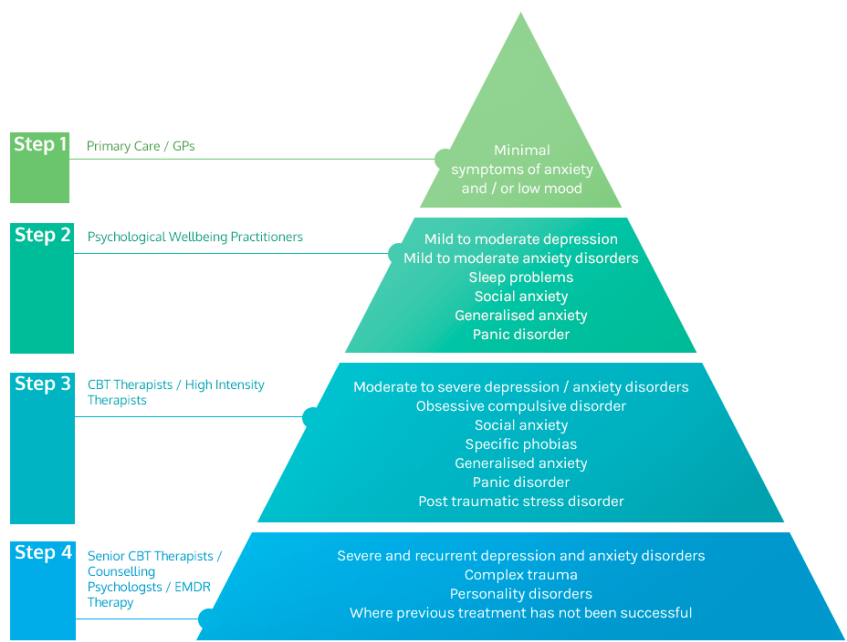

*Figure 1. Stepped care model used by IAPT services [MHM 2020].*

For the specific NHS services from whom we recruited PWPs for the purposes of the research reported here, the following on-boarding steps of a new client are described. Initially, clients undergo a face-to-face assessment to evaluate their suitability for step 2/ IAPT service care. If suitable, and the client has chosen SilverCloud as their treatment option, they are set-up with an online account to the programme that reflects their specific mental health condition (e.g., Space from Depression and Anxiety). The PWP then encourages the client to login to the service to familiarize themselves with the program prior to a first treatment session. This first treatment session, called TS1, is typically conducted over the telephone and serves to clarify: (i) the clients symptoms, goals and expectations of treatment; (ii) the set-up and use of the online program, including what tools are available to the client and how they can leave messages for their supporter (the PWP); and (iii) what is expected of the client in the coming weeks, such as the proposal to complete the first 1-2 modules, and explanations of their progress being reviewed weekly or by-weekly. Further, in keeping with a patient-centred care approach[2], clients are given a choice whether they would like their supporter reviews to be a *telephone conversation* or to receive feedback through *written communications online*. For telephone reviews, PWPs commonly allocate 30 minutes in their schedule, whereas online reviews are completed within a 15-minute time slot. The provision of telephone reviews was reported to be most common in this service. Only very few users were *instant* or *direct access clients*, who would come onto the service without a prior TS1 call.

---

[1] https://www.nice.org.uk/
[2] A patient-centred care approach suggest that treatment should take into account patients' needs and preferences, meaning that clients should have the opportunity to make informed decisions about their care in partnership with healthcare professionals [13].

*Figure 2. Example of the Supporter Interface in SilverCloud Health. It only includes data of 'fake patients'.*

Following access, all interactions between a PWP and their clients are coordinated via a specific Intervention Management Site. The site provides a quick overview that lists their current client cohort and indicates for each person whether they are due for a review and how much they have used the program via a simple traffic light indicator design (i.e., red – clients who did not log in; yellow – clients who logged in, but did not engage; green – clients who engaged). This enables PWPs to effectively prioritize who they need to reach out to the most, or most urgently (Figure 2).

*Figure 3. Example of a client review page.*

To review clients' progress, supporters can access information about clients both *directly* – via user messages sent to the supporter through the online intervention; and *indirectly* – via usage metrics of the frequency of clinical *questionnaires* completed, *messages* sent, *pages* viewed, *tools* used, and system *logins* (see Figure 3). Each of these summary items can be expanded to receive more specific details about what treatment contents a client looked at, and their specific responses in completing interactive activities (called *tools*).

Following a review of these data items, PWPs leave a feedback message for their client. For this, they can select and adapt messaging templates or specific text-paragraphs from a drop-down menu or write all text from scratch. The available templates are written by the supporters in their own words and are tailored to each client. Finally, PWPs can 'bookmark' existing or 'unlock' additional contents to suggest to the client specific areas of the program to look at for advancing user engagement and progress; and they can add any additional clinical questionnaires that they wish for the client to compete. Finally, they select a subsequent review date.

What remains less well understood so far are the processes by which supporters make use of the available client information that are provided to them via the interface in how they build up an understanding of their clients' situation, and how they construct their feedback. For example, how do they currently identify whether their clients encounter any difficulties with the program; have set helpful goals for themselves; or show any repetitive patterns? Which information do they consider as most important in assessing a clients' circumstances; where are potentially areas of uncertainty or knowledge gaps; and what might be challenges in how supporters choose what contents or activities to recommend or unlock additionally? Seeking to gain a better understanding of supporters' information needs and practices, the aim is to open-up the design space for how applications of ML could benefit their support practices.

## 3. Understanding iCBT Supporter Practices & Data Needs

The aim of this interview study is to learn about the processes by which iCBT supporters come to understand their clients' behaviors from the data that is available through the online intervention; and how they use this information to construct their feedback. This serves to identify opportunities for how ML approaches could be usefully applied to enhance supporter's existing work practices.

*Table 1. Participants role title and level of expertise in supporting clients via SilverCloud; the amount of years they have been using the service, and the number of clients currently assigned to them.*

| Current Role | Supporter expertise | Years using SilverCloud | Client load |
|---|---|---|---|
| Trainee PWP | Intermediate/ Expert | 1-2 | 4-5 |
| Trainee PWP | Intermediate/ Expert | 1-2 | 30 or more |
| PWP | Intermediate | 1-2 | 15-30 |
| PWP | Intermediate | 2-4 | 30 or more |
| PWP | Intermediate/ Expert | 2-4 | 15-30 |
| PWP | Novice | 1-2 | 15-30 |
| PWP | Intermediate/ Expert | 1-2 | 30 or more |
| Senior PWP | Intermediate/ Expert | 4-5 | 5-15 |
| Senior PWP | Intermediate/ Expert | 4-5 | 15-30 |
| PWP Clinical Lead | Intermediate/ Expert | 2-4 | 15-30 |
| PWP Clinical Lead | Intermediate/ Expert | 2-4 | 5-15 |
| PWP Team Lead | Intermediate/ Expert | 5 or more | 5-15 |
| PWP Team Lead | Expert | 4-5 | 15-30 |
| PWP Team Lead | Intermediate/ Expert | 5 or more | 4-5 |
| Step 2 Innovation & Service Improvement within Trust | Expert | 5 or more | 4-5 |

### 3.1 Participants & Procedures

We recruited 15 Psychological Wellbeing Practitioners (PWPs) from an NHS organization in the UK, all of whom regularly acted as SilverCloud supporters as part of their role. Our interview sample predominantly self-reported as female (1 male) and generally included fully certified PWPs who were very experienced at using SilverCloud to support their clients (see Table 1). Asked to rate their own level of experience as a SilverCloud supporter, the majority (n = 12) self-identified as "intermediate/ expert" or "expert". Participants further described having used the service a minimum of 1-2 years; some even reporting more than 5 years of use. Table 1 further shows the case load of each PWP at the time of the interview, which most often included 15-30, or more SilverCloud clients. For this particular NHS service, these client numbers typically represent ~50% of a PWP's overall case load.

The interviews covered three main topics of interest. They served to help the research team gain a better understanding of supporters' current: (i) *information review and response process* by asking open-ended questions such as "Can you describe to me how you are supporting your clients?" or "What do you do when you are giving feedback to a client?". We further inquired about the (ii) *importance and mechanisms of personalizing feedback* through questions such as "When writing your feedback, how important is it to personalize your message?"; or "Do you feel that you need to get a sense of the person to support them?". Lastly, we wanted to identify any (iii) *additional information needs that supporters might have* by asking for example: "Is there any information that would help you to support your clients better that is currently not available to you?", or "Has there very been a time, where you felt uncertain about your response to a client?". See Appendix 3 for the full Interview Guide.

All interviews were conducted remotely via video conferencing software (Microsoft Teams), and in one case via a phone call. Each interview lasted 1 hour and was audio recorded. All participants received a £30 gift voucher to a retail store of their choice to compensate them for their time spend in contributing to the research. The research study was carefully reviewed and monitored for compliance and privacy regulations; and approved by the NHS Health Research Authority (HRA, reference: 19/LO/1525). Each participant has been given a unique identification number to protect their anonymity, which are reported as P1-P15.

## 3.2 Data Analysis

All audio recorded interviews were fully transcribed and subjected to Thematic Analysis [3]. This involved an intensive familiarization with the data, and the identification of, and system search for, reoccurring themes in the data that were coded and developed in higher-level categories. Our analysis was guided by our main research question: What are the strategies and challenges of iCBT supporters for providing effective, personalized feedback to their mental health clients? Next, we present the six key themes that emerged from this analysis. For each of the themes we outline opportunities for ML.

The themes describe the importance for, as well as strategies and challenges of, supporters in:

(i) *understanding their clients' mental health problems and progress, as well as risks and associated blockers* for improvement that may need to be addressed;
(ii) *assessing and responding to how clients engage with the iCBT programme* to provide targeted support (*e.g.*, for low, irregular use, or unexplained changes in engagement);
(iii) *evaluating client progress with the programme* for deciding next treatment steps as well as identifying *struggles with the program content* that require more explanation and support;
(iv) *extracting and responding to relevant client information under time constraints* to maximize the benefits of support and minimize risks of overlooking issues that need to be addressed;
(v) *gaining a 'sense of the person' and forming a therapeutic alliance* to achieve a more personal connection, and through this, improve client engagement and therapeutic outcomes;
(vi) *developing the skills and confidence in effectively communicating with their clients*, especially if supporters are novices in guiding clients through the use of an online iCBT programme.

# 4. Challenges & Strategies for Effective, Personalized iCBT Support

To build up a picture and understanding of the clients' situation, supporters bring together a multitude of information that allows them to form assumptions about their client; how well they are working with the program; what struggles they may encounter; how they are progressing in their mental health; and how best to provide helpful guidance in response. In the following, we describe these work practices in more detail as well as the challenges that PWPs frequently encounter and the strategies that they apply to provide effective, personalized support to their clients. Our findings are organized according to the six themes that we identified.

## 4.1. Understanding Client Mental Health Problems, Progress & Risks

A key focus in the review practices of PWPs is to gain a sufficient understanding of their clients' mental health. Primarily, this includes assessments of: (i) the client's *main mental health problem* to be able to *identify the right treatment*, and *effectively focus* or *adapt the treatment* to the person's needs; (ii) how the client is *progressing in their mental health* throughout treatment, whether there are any *barriers to improvement in their mental health* that should be addressed, or whether there are any *indicators of risk* that have to be responded to immediately. Further, such assessments are not solely done by considering a person's clinical scores alone, but *carefully evaluated in the context of other information* that the PWP might have about a client and their life circumstances; and they are regularly discussed with other clinicians as part of case management supervision.

### 4.1.1. The clients' main mental health problem & how it determines the treatment focus

Initial screening assessments of the client and their conversation with their PWP during the initial treatment call (TS1) typically provide the supporter with key information about the person and their mental health. Such information is often recorded on a *client boarding card* within the IAPT system that provides basic health and demographic details about the person such as their name, date of birth, NHS number; as well as how many episodes and what types of treatment the person has had with the service; and any other treatment relevant 'labels' such as if the person has a long-term condition (LTC), or suffers from Post-Natal Depression and Anxiety (PNDA). It also includes a rating whereby the person can be ranked as 'red' – which indicates the need for support by more experienced PWPs – or 'green' for less complex cases, which guides resource assignments.

Conversations with the client during TS1 further enable PWPs to gain a sense of the person through descriptions of their struggles 'in their own words (P14)'. Here, they tend to pick up, for example, on descriptions of what the person has started or stopped doing in the past (i.e., hobbies, socialising behavior), or how 'psychologically minded' the person might be. For example, if a client demonstrates the ability to tease apart emotions from their thoughts or behaviours more easily, this can give the PWP an indication that they might be able to move forward with the treatment contents at a faster pace, or otherwise might require more time to do more awareness work.

In addition, PWPs described creating a ***problem descriptor statement*** in TS1. This statement describes the client's main struggle (e.g., health anxiety) in one sentence and serves as useful orientation in guiding their clients through the treatment. For this, the PWP asks the client to complete a series of sentence starters: *"'my main problem is…', 'this is triggered by…', 'I think that I feel…', 'I've started…', 'I've stopped…', and 'this has impacted on…'"* (P9). About developing the problem descriptor statement, P12 expands:

> *"It encompasses all of the thought, feeling and behaviour, physical symptoms and then we have one paragraph that clearly outlines and defines the main difficulty that you're having. So then I'll do a starter sentence to say I want to start a sentence, and then can you finish it off for me. So I would say, for example, my main problem is and then I want you to finish what you believe that looks like and then I often have thoughts such as and then they'll finish the sentence. And then we reflect it back and check that that's consistent with how they're feeling and if it captures the main problem in a paragraph nicely for them, and obviously 95% of the time it does and then if it doesn't, we just amend it as appropriate, really."*

As completing these sentences can be challenging, this is typically achieved collaboratively in conversation with the supporter. P13 explains:

> *"Yeah, so I guess it's a balance of what they think and what I think, just because a lot of the time clients say 'Oh, yeah, it's this' and I'm really not thinking that that's the case. So, it has to be a collaborative thing. But a lot of the time it might be a case of going through what they think and then seeing that actually maybe not that and then looking at something later so they've experienced it rather than they've not done it because I've told them to".*

At times, the problem descriptor may not accurately represent the core problem of the client (P1, P5, P13); it is only through the clients' later program engagement that supporters identify other underlying problems that suggest the client may not be accessing the right treatment for their needs. For example, whilst often assigned to the more generic 'depression and anxiety' program, especially in case of online self-referral to SilverCloud, clients may, at a later stage, describe experiences of trauma, abuse, relationship breakdowns and panic attacks for which CBT is not the most suited therapy (P1, P5, P6, P7, P11); or they show more specific symptoms of health anxiety, insomnia or depression, which can indicate that a different SilverCloud program might be better suited and more relevant to them (P5, P6, P12).

### 4.1.2. Client' mental health progress, mental health risks, and barriers to improvement

The vast majority of PWPs (P1, P2, P3, P4, P6, P7, P8, P9, P11, P12, P14) described that assessing their clients' current mental health state, how it may have changed across review periods, and especially if there were any indicators of increased mental health risk as one of the first and most important steps in their review process. To this end, they would review the clients' current and past clinical scores to assess any changes (P7, P13, P15) as well as their mental health risk.

A *mental health risk* is typically indicated to the PWP through a 'risk tab' within their interface that flags-up when a client is 'scoring higher on the PHQ9, question nine' item (P8), which asks about thoughts the person might have about self-harm and intend to end their life. In case of an increased risk, the PWP would often take action by giving the client a call (P1, P4, P6, P8, P11, P13) to assess what is going on for them as well as to clarify if their current scores actually reflects how they are feeling (P2, P3). P1 describes:

> *"Also obviously looking at the PHQ-9 as well as the risk tab, looking if risk has changed at all. If the risk has changed, I would usually call that patient. So, in my previous service, if risk was zero to one, wouldn't necessarily be concerned, not that that's an indicator that things are okay, but that was sort of 'our' protocol. In this new service, if there's a change of risk, just call the patient just to check in, because sometimes they may just need a voice, or they may just need to say 'Oh no, everything is fine. I just felt a little bit more low this week'. So, I think it's just safety is so important, and monitoring that particularly in an online platform is crucial."*

In few cases, the supporters also described to having picked-up on mental health difficulties and risk indicators in client messages (P6) and their engagement patterns with the program (P1, P5, P7, P8, P11). In the following example, P6 describes how both a person's description of having increased or decreased their alcohol consumption can be an important risk factor to consider:

> *"Obviously, that's something I always look at as well in detail is their risk if it has flagged up that they've filled in the risk questions. So if they've got thoughts of ending their life and they've stopped drinking, that can potentially – not always – but can potentially be because they're making themselves a bit more alert ready to do that. That's something I'd always bear in mind. And drinking more can be a depressant, alcohol can make you anxious, so that would be then something that I'd be putting about the psychoeducation in the programme to say, "It's understandable. Lots of people do it. You have a rubbish day, 'I'll have a glass of wine.' But actually, that can make you feel worse." Yes, it just gives me a chance to explore it, or at least bear in mind for potential risk factors as well."*

Similarly, P1 describes how changes in client behaviour: the absence of any messages to the supporter in combination with lower engagement with the programme and an increase in PHQ-9 score can raise concerns about mental health risk:

> *"I think for me it was actually a lack of comment with this patient. So, they had been logging in normally and then one week there was only one login, look at one page. And their questionnaire, their PHQ-9 had gone up as well as their low mood, as well as their risk, sorry. And for me I was just like… again, you know, it wasn't a drastic increase in PHQ-9, but I just thought 'I wonder what's going on here'. I sent a message 'I noticed that you perhaps haven't looked at as much and you haven't left me any comments. How are things looking for you?' And obviously because the risk went up, I thought 'I should call this patient' and actually I spoke to the patient, and the patient has been really struggling. Didn't necessarily mean there was any severe risk to the patient, but the instance of suicidal thoughts had increased and become more intense. So, it was helpful to kind of notice that change of activity and actually say 'Oh, what is going on here?'"*

For assessing clients' **mental health progress**, PWPs review <u>trends in the clients' clinical scores</u>. If those scores show a gradual improvement, the supporter won't spend much time reviewing these any further (P6, P15). However, if the scores remain unchanged; tend to go up and down; change suddenly or significantly in either direction (P1, P8, P14); or indicate a decline in mental health, then the supporter would take further action.

For example, if the client scores are not improving or are not where the supporter expected them to be, they will look for potentially 'barriers' (P2, P3, P6). This can include the following strategies: (i) asking the client in an online message or call about how they are feeling to better understand their circumstances (P3, P4); (ii) taking a closer look at key aspects in the questionnaire items on which the client may still score highly (P1, P6) that could be indicative for example of 'persistent sleep difficulties' or 'high levels of anxiety', which can then be responded to in a more targeted way; as well as (iii) considering other context information that would provide the base for further inspection. P6 for instances evaluates the still high client score in GAD in the context of that person's progress in the treatment journey so far:

> *"Because if they're on the depression and anxiety programme, but I know they're more GAD, and I know they've not got to the worry module yet, then I wouldn't be panicking too much if their GAD was high. I'd just think, "Okay. You're coming onto that next, so I'll recommend that now." Yes, a bit of a mixture of everything."*

In cases where the client does not improve after several weeks, they are brought to case management supervision (CMS) to re-assess the suitability of the current treatment approach for them (P11, P14, P15). In general, all clients are regularly reviewed in CMS, typically: when they are *new* to the service; *four weeks* into an intervention; had *no changes in their scores*; or showed *risks*. In CMS, supporters discuss the clients' main mental health problem and current treatment plan and consider their level of program engagement and clinical scores to identify risks of drop-out and strategies to address any treatment struggles (i.e., by changing the intervention, or stepping the person up to a more intensive care approach – see Figure 1). Describing the role of CMS, P14:

> *"Yeah, so it's case management, working through everybody they have on their caseload on a regular basis. It's to monitor, it's to make sure that everybody is discussed throughout the course of the treatment so that everybody, you can check the treatment they're receiving that it is appropriate for their presentation and check engagement and you can offer additional treatment or alternative treatments if they need something different. And you can monitor risk and like disengagements and you can appropriately finish somebody in the service as well. Yeah, so perhaps the, yeah, so you would definitely be looking at non-engagements on SilverCloud on there, and you would be trying to make a bit of a plan about what the person is going to do to try and engage the person again, which is usually just to try and speak to them and to decide if SilverCloud is still appropriate for them."*

It is important to note, that some supporters also described the need for caution in interpretations of the clinical scores as they may not exactly reflect what is going on for the person (P3, P6, P8). While frequent clinical assessments can provide a numerical indicator and trend of that person's mental health progression, these should not be given too much weight or consideration in isolation. Supporters therefore described efforts to contextualize the scores with information provided in client messages and conversation as well as considerations of their program engagement (*i.e.*, clients who tend to rush through the treatment programme in short time may be particularly 'anxious' – P7, P8), descriptions of physical symptoms, life circumstances and treatment fit. P8 for example describes how reductions in clinical scores may not mean the person is actually feeling better, making them reluctant to give their clients feedback about their scores:

> *"Well, it depends. If they've come down, because the scores are arbitrary. I know that that's quite a controversial point. I think that you can go down a dangerous route when you tell people, especially when they've improved, "Oh, you've improved now so everything is great" where that might not reflect exactly what's going on with them."*

Thus, for the interpretation of unchanged or potentially worsened mental health scores, supporters rely on a dialogue with their clients to clarify any interpretations. To this end, they find it useful to bring in other additional information, such as whether there have been any changes in the clients *personal life* or in their *type of employment* (P2, P6, P9, P11). Such information can help contextualise why someone might be struggling with stress and worries (e.g. exam times for students), or show heightened levels of anxiety (e.g., teachers lacking structure in their lives during term break), and can clarify if someone has easily the means to commit suicide (e.g., isolated farmer). All this enables supporters to adjust their assessment:

> *"(…) for example, it may be helpful in the instant access clients to know their occupation because not only does that serve purpose of what stress they may under, that also gives us an understanding of what access to means they may have if they were having suicidal thoughts. (…). So say for example, if they were a farmer, we're going to be like okay, so you may be isolated. You maybe have access to means first, ending your life, suicidal thoughts, self-harming. You start to be like okay, teacher, okay so your anxiety, your low mood maybe associated external from term time. We know that a lot. We, or I've seen it happen a lot that there's been a low in mood when they haven't got that kind of structure of term and what to do at work."* (P9)

### 4.1.3. Summary & Opportunities for ML

In summary, this section described how PWPs often gain an important first understanding of their clients' main mental health problems, life circumstances, and readiness for treatment through an initial assessment call. Such information, together with clients' programme engagement, serve as contextual resources for supporters to evaluate clients' mental health progression and guide their treatment. A key focus in assessing client mental health progress involves the regular review of clinical scores and their changes over time to detect any *mental health risks*, potential *barriers to improvement* (i.e., persistent sleep difficulties), as well as *requirements to re-evaluate the clients' suitability for a particular treatment* and associated needs for *changes in care* (i.e., switch to a different treatment programme; switch to telephone reviews; extension of PWP reviews; step-up in care).

Responding to these tasks, ML research might explore some of the following questions and opportunities:

- How could ML support early or automatic detection of 'mental health risks'?
- How could ML assist PWPs' understanding of why a clients' mental health might not be improving?
    - How could ML support the identification of 'barriers to client improvement'? This could help guide supporters in the detection of specific blockers and enable more targeted support? Such identification may be based on clients' clinical scores (and their language) relating to difficulties with sleep, relationship, or self-esteem; or be brought forward through 'significant increases' or changes' in questionnaire item scores.
    - What mental health difficulties (i.e., reflected through questionnaire item scores) may be particularly salient in predicting good or poor mental health outcomes? This could help PWPs prioritize which (potentially most) pressing mental health barriers to focus on in the review.
- How could ML help identify early if specific clients are likely to benefit from treatment, or not? This could help inform clinical decision making and treatment adjustments to ensure the client gets the 'right care' as early as possible and avoid wasting service/ supporter resources.

## 4.2. Assessing & Responding to Client Engagement with the iCBT Program

Alongside building an understanding of a client's mental health state and overall progress, PWPs also assess their general engagement with the treatment program. For this, they not only examine and compare across review periods how often a client logs into the service or the extent and depth of their engagement with content and tools; they also consider client communications and how their level of engagement relates to their clinical scores. In reviewing client engagement, supporters described: (i) the *characteristics of common program engagement patterns* that they commonly notice; (ii) *indicators of engagement struggles;* and (iii) what *feedback strategies* they employ to address any struggles, or promote more client engagement.

### 4.2.1. Patterns in client engagement and how they relate to client improvement

Asked to describe how supporters evaluate a client as 'engaging well' or 'less well' with the program, they described characteristics of different engagement patterns that they tend to notice over time. P14 explains:

> *"I think, yes, you start to notice patterns and types of people that will be the people that use it a lot versus the people who really like to share things on there, people who really engage with the exercises or then the people that are generally much better at just reading and digesting information. So, I think I would usually have in my own mind more of a mental representation of someone who is perhaps a little bit quieter, doesn't really feel the need to share lots of information, but they're still benefiting out of just gaining something through it. As long as they're still always logging in, working through material, I still would imagine them to be finding it some help. But I think it's harder to frame those people when you get less from them."*

In our analysis, we summarized and labelled various descriptions of client engagement patterns as follows: *Good Engagers, Content Rushers, Minimal Engagers, Little-progress-but-tries Engagers,* and *Non-engagers*.

***Good Engagers*** were described to neither do too much, nor too little on the programme (P7, P8), which means they would: login regularly; review 1-2 modules per week or review period; complete the various activities within those modules; and their completion of the clinical questionnaires would show a decrease in mental health symptoms. Especially the *coming down of clinical scores* suggests to PWPs that the 'treatment is working' (P13) and that the client is "using the program well and [that] its' helping them" (P11). The *use of tools* further signals that the client is "*doing the treatment, which is a positive, so you actually see them engaging with the material they're using*" (P14). About the relevance of tool use, P1 explains:

> *"For me, it's actually the tools. Because when you see that person has engaged with the tool, that's not to say they just looked at pages and didn't gain anything at all, that's not what I'm saying, but for me if I see that they've looked at tools, a lot of the time they will tend to repeatedly use those tools which gives me a really good indicator of how things are going. (…) I think tools to me is the give-away to see what that person has actually been doing. Again, that's maybe a bit of a sweeping assumption that because they've looked at the tools, they've gained something, but I suppose when you're working online that's your best indicator. For me, anyway."*

Additional indicators of good engagement are the extent to which the client 'engages in dialogue with the supporter' (e.g., by actively using the messaging function – P10); 'takes the time do things' (P2) by practicing 'the learned techniques rather than rushing through the program' (P9), and demonstrates 'good insight into the application of the program to their life' (P10). However, there is no one best, prescribed way of what constitutes good engagement and assessments of clients use of the programme are evaluated in the context of each person's circumstances and how far along in treatment they are. P13 describes, for example, changes in expectations of content coverage as follows, indicating the importance of identifying what in the treatment appears to be working for the individual and to focus on that:

> *"It does become less structured like that, the further you go along into treatment, because they find what's relevant for them, so it's more just really focusing on what modules they found the most helpful. So, we might spend two whole weeks on one particular part of one particular module once they've figured out that's quite relevant for them. So that does kind of change. We set it up as two modules between each review, but it can change as we go along."*

In addition to Good Engagers, a number of PWPs described the common occurrence of **Content Rushers**: clients who cover a lot of content in relative short amounts of time or few sessions, which is assumed to be overwhelming and unhelpful for learning. Examples include clients who reviewed the majority or entire treatment program 'in one go', 'within 1-2 logins' or within 'one or two weeks' (P4, P7, P8, P9, P10) as well as indications in client login and content view data that the person has looked at '50 pages' (P6) or '3-4 modules' (P13, P14) within a few sessions during a review period. P9 explains how such program use is unhelpful and indicates a need for more client guidance:

> *"(…) you'll find a client who comes into the treatment who's completed the majority of the programme, then you're like whoa, whoa, whoa, whoa. As helpful as it is to have the whole programme available to them, it can be quite overwhelming and then immediately like, well 'I've learnt nothing, I've done nothing different, I feel no different, I feel no different' and well that's because okay, 'yes, I understand where you're coming from, but you have just read through the whole programme', what's that taken you? A day? And this is the point of guidance, this is the point of it the self-help here. That's the self-help but we're guiding you through it, guiding you through the process. So as helpful as it is for them to have everything available to them, it can be a little bit daunting and overwhelming. So to rein that in a little bit would be really quite helpful."*

PWPs further described **Minimal Engagers**, who may login to the program, read content and complete the questionnaires, but are not actively engaging with the tools or leaving any messages (P2, P4, P7, P14). For many supporters the non-completion of the tool means that even though a client had gone through the contents, they are not 'putting the learnings into practice" (P7), which can be "a bit of a red flag that maybe they need a bit more support" (P2).

**Little-progress-but-tries Engagers** are clients who login a couple of times within a review period, but do not tend to make much 'useful' progress with the program (e.g., they only look at 1 content page). To supporters, this at least indicates to them that the client is 'attending the sessions' (P10) and 'trying' to engage with the program (P2); providing some insight into their attitude towards the treatment (P8) and general willingness to learn (P9). Client performances of intent to engage are especially apparent in the following example by P13:

> *"I've had a lot of people who will log in on the day of our appointment just so they can say they have been doing stuff. Which I always find quite interesting, because I can still see when they've logged in. So I do make a point of saying that right at the start of our treatment, that I can see when you've logged in and I can see how much you've done, and it is something that I will keep an eye on, so that they are kind of expecting me to pull them up on it later if they are going to tell me they're doing it but they're not."*

Lastly, **Non-Engagers** are clients who have not logged in, nor completed any of the clinical questionnaires during a review period (P4, P7, P8, P15). These clients become recorded as so called DNAs: Did Not Attend. Currently, supporters only find out about DNAs at the day of the schedule review (P10). In the absence of any client engagement, and thus data about the person, supporters cannot provide any specific feedback or tailored guidance. In those instances, PWPs tend to send a templated feedback message (see more about template use in Section 4.4.4). There's a number of additional consequences to a lack of user engagement:

Firstly, supporters described 'user disengagement' as a key 'struggle' to, and one of the most 'frustrating' and 'least liked' aspects of, their work (P1, P2, P5, P6, P13) that affects about 25-30% of their clients (P5). Repeated efforts to (re)engage clients can feel like hard work (P2), and, as a result, it can be difficult for supporters to remain positive and encouraging (P13). Thus, they expressed a desire to better understand 'the barriers for why clients may not login or do more of the program' (P13). Secondly, DNAs events present a missed opportunity for supporters to re-engage the client prior to the review date. This not only also complicates supporters' ability to manage their review time most effectively, it can increase risks of client drop-out, or risk causing their discharge from the treatment service if DNA events occur for two consecutive review periods. As additional causes for client drop-out, PWPs mentioned *mismatched expectations* of treatment (i.e., that therapy happens during telephone conversations rather than the online program – P4, P9); *preferences for face-to-face* treatment (P6, P11); *difficulties in understanding or using an online program* (P4, P5, P11) – especially by older adults or people with diverse ethnic backgrounds; and a *general lack of motivation* due to their mental health (P2, P3, P12).

Further to these concerns on non-engagement and drop-out, PWPs suggested two additional areas for improvement where more or different data representations could facilitate their understanding of client engagement:

Firstly, to be able to better assess depth or quality of engagement, a number of supporters described the limitations of only having access to data describing the frequency of logins, content views and tool uses, but no information on the duration of engagement (P6, P13, P14). P13 describes how having a <u>clearer understanding of the timeframe</u> can assist in interpretations of client engagement:

> *"I guess maybe time. I know they're logging in and I know they're looking at five pages, but I don't know if they've spent two minutes on that or two hours. (…) Whereas if I knew how long they were spending… because someone might only log in once, but if they've spent an hour, that's a good chunk of time to be spending on the programme. But because it only says they've logged in once, it could lead me to think they're not really engaging with it. So, I guess a timeframe might be quite helpful. Even if it's over the whole two weeks, so 'Oh, you've spent two hours on the programme in the last two weeks' or 'You've spent two minutes on the programme in the last two weeks', just a bit of a guideline. And I think it would be helpful for them as well to keep track of how much they're engaging with it."*

Secondly, one supporter (P1) described the need to <u>facilitate assessments of 'engagement trends' over time</u>. While the current interface (Figure 3) gives an overview of a clients usage of the programme since their last review, it can be harder to compare those statistics across review periods to identify trends. P1 describes:

> *"So, what I, this is where things get difficult. So, I either have to open up two windows and looks and it side by side, or I can get up my IAPT notes, shrink both windows and then compare it that way. Because on my IAPT notes I will have, from the previous review, I will have put up 'This is how many pages they looked at, these are what the pages are' and then I can do that comparison next to that week's SilverCloud review, if that makes sense. So, it's kind of just a bit of window juggling. Not very technical, but that's how I compare. (…) If there was some kind of almost like a mini summary table or something next to… so you know when you go onto someone's page, as you're about to write the review it shows you they looked at 15 pages, they've completed five questionnaires, those, if there was almost like underneath just a summary of what they looked at previously then that would be amazing, just the session before. Because then you're not doing as much juggling and you can, if you're opening them both up and you can compare it that way, that would be incredible. That's not a very technical way of describing it. If you could see them literally one under the other, the numbers, obviously very clearly 'This was this week, this was the last review' that would be amazing."*

### 4.2.2. Indicators of engagement struggles

Temporal comparisons that could support assessments of engagement patterns can also help in identifying *client engagement struggles* that supporters would need to respond to. This includes <u>noticing changes in client activity over time</u> (P1, P8, P15), whereby client engagement in some weeks, but not others, can foreground difficulties. P15 describes frequent observations of such engagement changes in university students:

> *"I think you can sometimes tell, like I see quite a lot of young people, like people at university and things, and you can tell when things have got chaotic for them. They might start off doing really well on the programme, and then they drop. Then you boost them, and then they drop. Then when you speak to them, it's because university stuff has taken over."*

*Non-completion of tools* can further indicate engagement difficulties with the programme as well as any other *consistent patterns of struggle* that may surface. P14, for instance, describes observing the trend that a client repeatedly starts, but not completes a content module:

> *"Versus, then the other side of things would be some way of teasing out, almost noticing things they're starting to find difficult. Whether it's a case of this person has left the last three modules incomplete or something that you can see 'Oh okay, this person is not actually getting on with that very well' or 'This person hasn't completed any exercises' that sort of thing."*

### 4.2.3. Feedback strategies for supporting client engagement

Supporters described a range of strategies that they employ in their feedback to either (i) support continued client engagement or (ii) aid client motivation to engage more, or more frequently with program contents as well as the supporter.

If a **client engages well**, PWPs would often start their message by *acknowledging their good level of engagement*, often by referring to or *summarizing the contents that the person had read* (P4, P5, P11, P12, P13, P15) and *providing assurance* that they have 'done well and should keep going' (P15). P13 explicates:

> *"So, once I've kind of seen what they've done, I would start by just leaving a general welcome of 'How have you been? It looks like you've been doing quite a lot, which is great' just that kind of encouragement if they've engaged quite a lot."*

In addition, PWP would also highlight: any *examples of client experiences* that they had described in comments or worked through using the program activities (P5) to encourage further sharing and tool use; as well as any *successes*, where the client indicated how aspects of the treatment were helpful to them (P1, P11, P12) to remind them of their progress and reinforce treatment directions that appear to be working for them. Table 2 summarizes examples of positive reinforcement strategies that PWPs apply to support client engagement.

If a **client had not engaged much**, supporters would again respond positively by acknowledging – however little – the client had done on the program, and then offer *reminders and tips* to help improve clients' engagement going forward (P3, P4, P5, P6, P15, P16). Table 3 summarizes examples of such reminders and tips for engagement that supporters described to encourage: more engagement generally; more frequent program use; more balanced/ effective treatment use; prioritization of program use; practicing; and the use of alternatives (e.g. other media) that may be more suited to the preferences of the person. To give an example:

> *"Great work. I noticed that you've looked at this module, but perhaps you might want to think about doing the program slightly more regularly for it to be as effective as it can be."* (P15)

Furthermore, if a client isn't leaving any messages or completing tools, PWPs would deliberately *ask questions* to encourage more interaction and dialogue with the person and promote tool use to be able to better assess client understanding and progress (P1, P2, P3, P4, P6, P10, P13, P14). Examples include: *"Oh, it looks like this was quite helpful, what did you think?"* (P13); *"I've looked through this and it seems like you found that module helpful. What was your favourite thing about it?"*; *"Does that feel reflective? Does that make sense? Have I understood properly?"* (P6).

*Table 2. Examples of Positive Reinforcement Strategies for Client Engagement.*

| Summarizing any progress client has made | "I might start by saying which is in line with the template, 'I hope you're well, well done for logging in three times since our last review, I can see that you've now completed X and X module, I hope you found these useful'." (P5) |
|---|---|
| Acknowledging/ reinforcing 'really good' engagement | "If they've done a really good engagement, I'd probably be encouraging them further and saying how well they've done maybe and then where to go next." (P4)<br><br>"'How have you been? It looks like you've been doing quite a lot, which is great' just that kind of encouragement if they've engaged quite a lot." (P13) |
| Highlighting positives/ client successes | "So I think if they've said something that is particularly positive about what's happened, I'll want to comment in on that (…)" (P12); "So, successes, anything the patient found was helpful." (P1) |
| Foregrounding client successes in learning & linking with interventions | "(…) they could say, I felt great on Tuesday, I managed to get up and I managed to do a little bit of housework and I met a friend for coffee. I'd really comment on that and say, how helpful that is and you need to hold onto the fact that how good you felt after you were doing the activity. So it's still linking it in with the intervention, the behavioural intervention, for example, and reinforcing the idea that when you do activity, you do feel better so how can we learn from that experience and kind of maintain that learning to help continue motivation." (P12) |
| Appreciating client sharing examples/ experiences | "So they'll say things like that and then I'll respond back and say, oh okay, well thank you for sharing that, that's really helpful to know, I'm pleased that you can see the association there and that's what we're looking to do to explore, something that may be helpful for you, maybe looking at the sleep programme." (P9)<br><br>"(…) my response back would be first of all just thanking them and congratulating them for doing the tool, because otherwise they won't do any tools next time if I tell them that it's wrong." (P13) |

*Table 3. Examples of Reminders and Tips for Improving Client Engagement.*

| More engagement generally: "The more you do the better" | "The more you put into it, the more you're gonna get out of it"; "the more you do the better, the more activities you complete and share with me and the more information you share with me, the more tailored my feedback can be and those sorts of things." (P5) |
|---|---|
| More frequent program use | "The more regularly you use it, the better the outcomes." (P15) |
| More balanced/ effective program use: "Little but often approach" | "(…) the way SilverCloud works very effectively is a 'little and often' approach" (P3)<br><br>"Doing a little bit often, so not just sitting for long periods, but doing a little bit often." (P15)<br><br>"We suggest a little and often approach. So if we schedule 10 to 15 minutes a day at the same time of each day and try and form a habit out of that, we're more likely to stick to those habits." (P4) |
| Prioritization of program use: Scheduling/ goal-setting/ getting started | "You're logging in a bit more. Why don't you set yourself a goal of logging in?" (P6)<br><br>"One of my favourite ones which a lot of people respond well to is scheduling in the time, like a GP appointment, and if we make a point of saying that it's a GP appointment then we're more likely to attend it because when we do book a GP appointment we don't really just not attend or just forget about it, we actually go to them because we're prioritising our health and our mental health is just as important as our physical health. I give those as tips." (P4)<br><br>"Have you tried it?" (P10) |
| Practicing the learnings | "Taking tools and putting them into your day to day life. Practising." (P15) |
| Using alternatives that work: Creating own templates | If client struggles with motivation to login to a computer, paper diaries and other means to record their mood could help (P4) |

Another strategy for sustaining and building on existing client engagement is to *follow-up on what client had previously said or done* (e.g., in the tools). For example, a client may have stopped with goals or scheduled activities that they defined (e.g., they only did it for the first 2 weeks), which creates an opportunity for the PWP to follow-up on these goals and proposed activities to help the client build-up that practice (P1, P3, P5).

If client engagement is low, supporter may also choose to *leave a message with their contact details or calls the person* to see what is going on for them and whether they are happy to continue with the program, or would like to switch to telephone review to avoid possible drop-out without any further contact (P3, P10, P15).

Finally, if a **client did not engage at all**, as described above, supporters wouldn't be able to provide a personalized review and instead would often opt to send a standard DNA template (P4, P5, P6, P7, P10, P13).

### 4.2.4. Summary & Opportunities for ML

In summary, this section described how PWPs develop an understanding of their clients engagement with the treatment by noticing specific patterns in their use of program contents, tools and interactions with the supporter; and relating those uses to assessments of mental health progress. Depending on identifications of different client engagement types, which we labelled as: *Good Engagers, Content Rushers, Minimal Engagers, Little-progress-but-tries Engagers,* and *Non-engagers*, PWPs would tailor their feedback to either (i) reinforce good engagement, or (ii) encourage more extensive or more frequent uses of the program, its tools, or the sharing of personal client experiences and struggles so as to increase overall client engagement, their ability to provide more tailored and effective assistance, and to reduce risks of client drop-out.

To better assess *client engagement patterns, trends* or *struggles*, supporters expressed the desire for having: (i) access to duration information and (ii) easier means to review client engagement over time (i.e., across review periods). Such temporal information and comparisons could help approximate *how deeply a client may review contents* (not just how many pages they clicked on) and support the *noticing of changes in client activity* (i.e., indicators of struggles). Furthermore, PWPs explained the (iii) limitations of providing effective feedback and of increased risks to continued client engagement, if not client drop-out, in cases where the person 'did not attend' (DNA) treatment at all during a review period.

Responding to these tasks and difficulties, ML research might explore the following questions and opportunities:

- How could ML support PWPs understanding of client engagement?
    - How could ML assist in the identification of 'engagement trends' as well as 'changes in behavioral patterns' that may be reflective of specific struggles?
    - What are opportunities for employing ML to learn about characteristics of good/ effective program uses vs. less beneficial uses?
    - How might this enable us to learn more about what (re-)engagement strategies might be particularly effective for different types of program engagers?
    - All these could enable supporters to more effectively tailor their feedback strategies.
- How could ML aid the earlier detection of a person's risk of 1$^{st}$/ 2$^{nd}$ DNA in the next review period? This could enable PWPs to reach out to clients earlier to support engagement or the sending of (semi-)automated notifications to reduce risks of drop-out, ensure better engagement as well as better resource and time planning for supporters (i.e., having sufficient client use information to deliver a review, or identifying key blockers to engagement earlier to adjust care).

## 4.3. Evaluating Progress & Struggles with Program Contents or Tools

In addition to understanding a client's mental health and engagement with the treatment, PWPs draw on the available client data to evaluate how well the person understands and utilizes the program contents and tools. In this section, we outline how different data sources provide insights to supporters about how a client progresses with the therapy content and tools. In particular, we describe (i) how PWPs extract from client messages, their engagement with tools and their use of language *information about client learning, or experiences of struggles*. Responding to gathered insights, we then point to (ii) a number of *strategies that supporters employ in tailoring their feedback* to ensure they *address any identified barriers* as well as how they *determine how best to move the client forward in their treatment journey*.

### 4.3.1. Indicators of client learning or struggles that suggest specific support needs

While information about how many times a client logged in and what content pages they viewed can be important indicators for engagement and what interventions a person looked at, such data is less relevant for understanding how well a client has taken in the programme contents (P1, P2, P13). For this, PWPs instead consult the contents that the client has shared with them. While supporters can derive insights about what program contents a client finds helpful or unhelpful from client 'bookmarks', 'repeated completions of a tool' (P10), or client 'ratings and evaluations of each module' (P2, P7, P9, P11, P13); they focus in particular on *any direct client messages, clients' use of language*, and their *responses in completing the program tools*.

In terms of content sharing, supporters described especially **client messages and journal entries** as one of the most effective ways for them to extract relevant information about a client and gain a sense of the person (P3, P5, P6, P14). Through leaving messages for their supporter, clients can 'directly tell' their PWP what is going on for them; what they are learning; how they are feeling; and whether they are encountering any struggles. Therefore, messages provide a clear pointer to 'what's important to the person' (P6); thereby inviting most opportunity for understanding client progress and struggles, and as a result, for providing relevant, personalised feedback (P8). Supporters therefore emphasized the importance of, and to always prioritize, check for, and respond to, client messages (P3, P4, P6, P9, P8, P9, P10, P11, P12, P14). P14 exemplifies this:

> *"I think the messages, whenever someone left a message, that was so valuable because they could really tell you so much more than what you were getting just by looking at it. The clients that just used it to just read pages and things and not really do, it's what the client puts on there that makes the biggest difference to being able to help them better. So, in their personal message, anything they would say to you."*

Implicit in any contents that clients share through the program are also cues provided through their **use of language**. Here, some of the supporters (P2, P6, P8, P11, P12, P15) referred to specific <u>language makers</u> that can indicate 'positive behavior change' as well as specific 'struggles'. P15 provides examples that reflect positivity and progress in client expressions: *'Today is a much better day. I'm really learning things. I'm changing things in my life.'; 'I used to do this, now I'm doing this.'; 'I have found this bit really helpful. I have used such-and-such.'* Similarly, language can also be expressive of: a client's emotions (e.g., *I'm sad', I'm angry', 'I'm worrying all the time'*); struggles to 'concentrate' or be 'motivated' (P11); and covey how the person perceives themself. For example, supporters described how language can indicate that a client might have "quite high expectations and perfectionism traits" (P12), which can show through a person repeatedly expressing what they 'should be doing better' or 'should be doing more of' (P8). In addition, struggles with low self-esteem can also often be noted through language expressions of self-blame. P12 explicates:

> *"For example, they might say, 'look it was my fault that this happened' or 'it's because I'm not good enough'. 'I think I'm failing', so a lot of the focus of attention is around them and them being the ones that are responsible or to blame, I guess, if things are maybe not working out, rather than there being something wasn't quite right at work, 'my targets were quite high for example and I did struggle but I did the best that I could'. That's more likely to be 'it's my fault because I'm a failure', 'I couldn't cope, and other people would be able to do it'."*

Finally, supporters described deriving important insights from **clients' tool uses** (P2, P3, P5, P9, P10, P13, P14). Here, depending on the treatment program that the person is enrolled in, PWPs mentioned especially the completion of tool such as: the Thoughts-Feelings-Behavior (TFB) cycle, Understanding My Situation, Worry Tree, Mood Monitor and Behavioral Activation (P2, P3, P5, P7, P9, P10, P12, P14) to offer rich opportunity for understanding and personalization through the provision of 'examples that are specific to the client' (P13), where they describe their situation 'in their own words' (P14), and that serve as evidence that they 'fully comprehend the taught interventions' (P6). Describing how completing tools and client feedback can be indicative of their competence with the programme and progress in learning (P1, P6, P7, P9, P10, P15), P9 shares an example of how uses of the TFB cycle enabled a client to have greater awareness of the relationship between their anxiety and difficulties to sleep:

> *"So something along the lines, so we have understanding feelings which allows them to explore their thoughts, feelings, behaviours and emotions. And within that, that's kind of like the bread and butter for CBT. So with the sleep, they may comment, oh I never noticed that sleep may be associated with my anxiety. I'm lying at night awake thinking about things which is impacting on my sleep. So they'll say things like that and then I'll respond back and say, oh okay, well thank you for sharing that, that's really helpful to know, I'm pleased that you can see the association there and that's what we're looking to do to explore, something that may be helpful for you, maybe looking at the sleep programme. (…)"*

Simultaneously, how clients complete the tools can also indicate areas where the client may have missed or misunderstood something, or where the tool use may not yet be optimal and could benefit from some additional explanations and guidance (P1, P2, P3, P4, P5, P6, P7, P8, P9, P12, P13, P15). For three tools in particular PWPs frequently noted client struggles: the *TFB cycle* (P4, P6, P9, P13), *Worry Tree* (P5, P7), and *Hierarchy of Fears* (P5, P9).

For the **TFB cycle**, supporters would identify that clients would confuse for example thoughts and feelings (e.g., the client puts down 'lonely' as a feeling, when it's a thought – P9); or they might miss to include key behaviors that they may have mentioned to the supporter at other times (e.g., their initial assessment), meaning that they are not getting 'the most out of that tool' (P13). P4 describes a checking of clients' TFB cycle can surface difficulties in client understanding that they then can clarify for them in their review feedback:

> *"There's no right or wrong way of using them I guess but there's maybe ways that they have used them that could be used differently. For example, there's a thoughts, feelings, behaviour cycle and sometimes when people fill that in they might get their feelings mixed up with their thoughts. So it's about maybe spotting those little mistakes and clarifying those a bit more."*

For the **Worry Tree**, supporters described how the wording of the tool led to confusion for clients in the past (P5, P7) as it suggests the differentiation of *real* vs. *hypothetical worries*. However, for clients, all their worries tend feel as being 'real'. As a result, supporters often notice how clients start using this tool to problem-solve 'hypothetical' worries. Explaining how this isn't helpful, P5:

> *"(…) on the programme there's a module for Managing Worry and there's something called a 'Worry Tree' which helps people decide what type of worry they're having. So it talks about real or practical worries which is where we worry about something that we can do something about so we might need to do some problem-solving to come up with the solution for it, or hypothetical worries which is when we worry about something that we have no control over so it tends to be future-based like 'what if?', so, what if when we go on holiday next year we miss our flight? Well, actually, there's nothing you can do about that right now."*

Lastly, supporters also described client struggles in the completion of the **Hierarchy of Fears**, a tool where clients are asked to develop a graded list of feared situations that they then start exposing themselves to (step-by-step) to reduce their fear. However, instead of describing different actual physical trigger situations, clients would often enter their thoughts, worries or fears (e.g., 'myself getting ill, my children getting ill, my husband getting ill' rather than 'reducing checking behaviours' such as taking their child's temperature everything they feel hot – P5); or struggle to describe graded steps towards exposing themselves to their feared object. Regarding the need to support clients in identifying useful exposure steps, P9 explicates:

> *"(…) it just doesn't make rational sense if, in terms of, from a therapeutic perspective, to actually achieve it. So we'd work that through with them and say, is this step achievable? So for example, I'm scared of a spider, so going from one step of viewing it on YouTube, as opposed to holding it, the next step like, is there any steps that we can have in between there that can allow you to expose to it greater."*

### 4.3.2 Strategies for tailoring support and advancing the client in their treatment journey

Based on supporters understanding of client progress or struggles with the therapy programme, they employ a number of strategies to provide tailored guidance for: (i) helping the client overcome any identified barriers; (ii) deepening their understanding of CBT; and (iii) moving them effectively forward in their treatment journey.

To help **overcome client struggles**, PWPs predominantly described two approaches: Firstly, if clients described difficulties in processing what can feel like an extensive amount of content and reading material to work through, supporters would often propose to 'break things down into smaller steps' (P8, P11); to take their time in working through the programme; and normalize any challenging situations (see further Section 4.5.3 on normalization strategies as part of patient-centric care). P11 exemplifies:

> *"Some people, when there's a lot of reading sometimes, they say that, 'I spent a long time going through that module because it was quite a big module.' So sometimes, when it's got a lot of the tools attached to it, as well. So, I would normalise it, saying, 'Yes, I understand. It's a really big module. Maybe take a little bit longer. Maybe take some breaks in between sessions and do smaller sessions on the program.'"*

Secondly, PWPs described 'providing alternative explanations' (P8) or 'additional learning materials' (P2, P3, P4, P5, P6, P8, P9, P11, P12, P14) to help clarify programme contents – such as how to distinguish thoughts from feelings – and assist clients in their tool uses. P12 provides the following example:

> *"'I feel like I can't cope, and I feel like I'm not doing good enough'. So those are actually thoughts, so it would be saying, okay, when you're thinking I'm not good enough and I can't cope, how does that leave you emotionally* *feeling and trying to attach their feeling there [in the tool] because otherwise we might fall into the problem of trying to challenge feelings, but then those aren't feelings. So those need to be moved into the thoughts section [of the tool] and then we need to identify what is the emotion that you get as a consequence of having that thought."*

If supporters identified a specific client struggle with the programme or mental health blocker, they often described providing additional psycho-education contents to help address the underlying issue. To this end, they would either copy extra materials into their feedback messages (P3); link to external resources (P2, P3, P5, P6, P7, P13, P15); or unlock additional content modules on the programme (P2, P3, P4, P5, P6, P8, P9, P11, P12).

For certain client difficulties such as problems with sleep, relaxation, self-esteem, relationships, or grief, supporters also have an array of additional programme modules available to them that they can 'unlock' for the person (P2, P3, P4, P5, P6, P8, P9, P11, P12, P14) to help them learn the skills that can overcome these blockers. P9 exemplifies:

*"So with the sleep, they may comment, oh I never noticed that sleep may be associated with my anxiety. I'm lying at night awake thinking about things which is impacting on my sleep. So they'll say things like that and then I'll respond back and say, oh okay, well thank you for sharing that, that's really helpful to know, I'm pleased that you can see the association there and that's what we're looking to do to explore, something that may be helpful for you, maybe looking at the sleep programme. So perhaps now, between now and our next appointment, use the sleep programme to help you manage your sleep better or to consider additional sleep techniques."*

PWPs also look for opportunities in client responses that enable tailored explanations that serve to ***clarify and extend the key learnings of CBT*** (P3, P5, P6, P8, P12). In reviewing how the client is applying what they read to their specific circumstances, supporters can use these as examples to further extract and emphasize core CBT insights to deepen understanding. This is illustrated in the above examples where supporters describe providing additional examples, explanations and reading materials to clarify, i.e., the relations between thoughts, feelings and behaviours; and also the following example by P5, who provides additional explanations to help their client realize how holding on to unhelpful thoughts or behaviours can maintain their problems:

*"(…) if we can get from them what they are thinking or what they are doing that might be unhelpful, we can then reflect back to them, 'you said in your cycle that when this happens you tend to do this, what we know is that that is an unhelpful behaviour and the way that that maintains the anxiety is by this…'.  So it's always helpful to highlight to people what it is that they're thinking or doing that is unhelpful that's maintaining their problem, because that's what they need to challenge and change and that can be very different for everyone."*

Furthermore, supporters described making recommendations to use specific tools or contents that the client might benefit to review next as a way to ***moving the person effectively forward in their treatment journey***. Such recommendations are tailored to the personal circumstances of each client (P5, P6, P9) and depend on their (i) treatment goal as well as (ii) engagement and progress with the therapy.

In terms of treatment goals, the specific SilverCloud program that the client had been assigned to following assessment – such as the 'Space from Depression and Anxiety' – already presents a tailored version of CBT for addressing comorbid symptoms of depression and anxiety. As a result, many supporters described how they generally tended to follow the structure of the given programme type in how they may scaffold specific content recommendations or suggest tool uses (P2, P4, P5, P10, P14). P4 explains:

*"Nine times out of ten I follow the structure of the programme but sometimes, the odd occasion, I would jump forward. It depends on the client's needs. I'd assess that maybe from the assessment notes or from that first call that we would have and see what they say and what their goals are really."*

Moreover, supporters described directing and prioritising their feedback and content recommendations in line with the clients' main problem descriptor (see Section 4.1.1 above) and what they want to achieve through treatment (P3, P4, P5, P6, P9, P10, P11, P12, P13). This prevents that the supporter 'goes off on tangents' (P5) and instead addresses the most 'relevant difficulties' and suggests contents of most importance for the person's presenting problem. P12 explains:

*"I think it really depends on the individual, so the most important thing for me is I try to refer to what our main problem statement was and what the goals for the client was and because that's important that we don't go off*

*track, which can be easily done. So by making sure that the goals that they've set aligns with what we're working towards will help me determine what I'm going to respond back."*

For choosing contents to recommend to clients next, supporters described to rely on their experiences (P2, P9) of what their clients find most helpful depending on their main problem area. As examples they highlighted: the 'hierarchy of fears tool' (P9) for clients with phobia; the 'boosting behaviours module' for people with depression; and the 'worry module' for clients with anxiety (P2) as key materials.

When providing these and other content recommendations, PWPs also tend to make an effort, in their feedback message, to clarify how previous learnings and upcoming treatment modules and tools relate, and are of importance to the clients' main problem (P10, P13). For instance, whilst a client may be on a mixed depression and anxiety program, anxiety might present their predominant problem. P13 describes how they specifically emphasize the alignment of the program contents with the person's main problem:

> *"So I guess if it's earlier on it would be things that would lead them quite nicely onto future modules, or things they might be coming across later. So if they have filled in an activity about depression focus, but we know from the assessment call that actually the anxiety is a lot more of a problem for them, I would want to make sure that I've picked up on that to link it onto the anxiety to make sure that it is still relevant and things they know are going to come up in the programme going forward."*

As described above, supporters assess from client comments and contents if the person understands the treatment materials. If the client shows good understanding and demonstrated sufficient skills practice, supporters consider them as 'ready to move forward' (P2, P3, P6, P14) in the program; or otherwise may adapt the pace to ensure adequate time, focus and learning of key aspects of the therapy. P3 exemplifies:

> *"But unless they've said that, "Yes, I understand this fully now. I've been implementing it. I'm now ready to move forward," that's the time that we do so. But unless that's happening, I tend to stick on what they need to focus on at that point. Otherwise, they're just going to be confused and it's an information overload, and they won't exactly know what we're doing already."* (P3)

One key strategy that supporters use to help advance their clients through the programme is to anchor existing client work on the programme to the next module or tool (P2, P7, P6, P9, P14). To do so, they often create 'bookmarks' for the client to guide their focus on what contents to look at next (P7, P13) and explain in their feedback how previous tool use and learnings act as building blocks for subsequent activities. P7 explains:

> *"So if they've done thoughts, feelings, behaviour cycle, I say 'Well I like this example of X, Y and Z that you've done, it's really interesting' and then I will link that to the next part in terms of creating an action plan. So I'll say 'You've done this example on your thoughts, feelings, behaviour cycle, this will be really useful when you go into challenge your thoughts [module and subsequent activities].'"*

Having described their **strategies in tailoring the program to progress the person**, three supporters expressed desire for occasionally more flexibility in how to put together the program and guide their clients (P6, P9, P10). Currently, supporters can only 'unlock' a set of additional modules for the specific programme that the person is using, but they cannot 'easily pick and choose modules from other programs' (P10). The following quote by P6 illustrates how having more control and flexibility in personalizing the program structure could help address issues with content pacing (P6, P9), whereby clients rush through materials that they do not want to skip to adhere to the pre-set order of contents:

> *"I think it might be a long-term plan, but having just a real mix of, "This is your personal programme," rather than, "You're on the space from GAD," I think would be really helpful. So you can actually go, "No, I think this module would be helpful for you." You have the basics, they have the introduction, they have the ending, but you choose all the different modules that might be in there and what order they come in. I think that would be good, because I think that can be where it's difficult is a lot of the time, clients want to go through the programme in the order it's there. So if you recommend looking at the fifth module, but they're only on the third, they'll often whizz through the fourth one to get to the fifth one. So I think just being able to pull upon any modules would be helpful, especially as I have had instances where they're on the Space from GAD, for example, but they need a bit more support. Then you have to just email them a booklet and then they're not really using the programme anymore."* (P6)

Having opportunities to personalize treatment was highlighted as important to be able to better respond to inter-individual differences in clients' mental health experiences (P2, P13). Describing how different programs and contents are required to better address the diverse symptoms and care needs of clients, P2 explicates:

> *"Because I suppose with, well like anything, everybody wants something personalised to them, but when it comes to healthcare it is so different for everybody that you can have standardised templates to a point, but again, everybody's depression is going to be different. Someone might have all the motivation in the world, but then negative thoughts are terrible. With anxiety it's called generalise anxiety because what I worry about, you might not worry about."*

Creating more flexibility and having more components that supporters can choose from for their clients – such as a broader range of modules, selection of relevant personal stories, and so forth (P6) – can expand their scope for personalization; and through this, may increase a **sense of supporter agency** in guiding their clients. Being able to take specific actions in tailoring the program has been particularly apparent in the way that supporters described to 'enjoy unlocking modules' (P8, P14); with some sharing that they 'always unlock the relaxation module for all of their clients' (P2), or unlock all available modules for their clients at the end of treatment to expand SilverCloud as a resource for them (P8). P8 states:

> *"You can advise them of different models that they could use and then also, like I love to unlock stuff as well, like I'm always unlocking everything for everybody. Like I try not to overwhelm them with stuff but definitely I like doing the core beliefs and the self-esteem modules. I like those."*

Simultaneously, supporters also described feelings of frustrations when clients did not adhere to their specific recommendations and difficulties to comprehend why that might have been the case (P2, P6, P11, P13). In this context, they reported limitations to the data that is available to them, which currently does not provide them with any feedback on whether a client has actually seen and read their message – and thus had awareness of the content recommendation – nor are their easy means available for them to 'check' whether homework was completed outside a manual back-and-forth between previous support message and newly covered content.

### 4.3.3 Summary & Opportunities for ML

In summary, this section explained how PWPs draw on available client data to evaluate how well the person understands and utilizes the program contents and tools. Much of the insights into client understanding are derived from client messages to their supporter where they directly express progress or struggles in learning. Clients' use of language can further reveal insights about their mental health state and behavior changes (i.e., indicators of positive change). A central focus of PWPs is the review of clients use of program tools, which are crucial for helping clients apply the content learnings to their situation and support the practice of core CBT skills. Building on their therapy expertise and familiarity with the SilverCloud tools, supporters look for any errors or sub-optimal use of the tools that they may need to address.

To help overcome identified misunderstandings and barriers to content or tool use, PWPs described employing a number of strategies in their review feedback to ensure clients achieve the desired benefits from treatment, such as: normalizing; the provision of alternative explanations; and additional learning materials – including external resources and the unlocking of specific psycho-educational add-on modules. Building on their treatment expertise, supporters also utilize client examples to emphasize core CBT learnings not just to clarify, but also deepen client understanding. To effectively advance a client through their treatment journey, PWPs make recommendations for specific tools or contents for clients to review next. Their choices are informed by each clients' main treatment goal; their pace and success in progressing through the program; as well as supporters own experience in what program components are particularly important for specific mental health conditions (i.e., the worry module for people with anxiety). PWPs further described explicit efforts in their feedback to anchor how existing client work or presenting problem links to upcoming activities so as to clarify the relevance of the therapy contents to clients and their progress through treatment. In some instances, supporters expressed the desire for more opportunities to personalize the structure of a treatment program to achieve a better tailoring to each person's unique treatment needs as well as to increase their sense of agency through more flexibility and control over the order and pacing of contents (especially via module unlockings).

Responding to these tasks and desires, ML research might explore the following questions and opportunities:

- How could ML (i.e., NLP) assist in detecting whether clients are demonstrating learning experiences and improvements to their mental health (i.e., from positive changes in their language) – see 'moments of change work' by Amit Sharma as example? Could we detect 'eureka moments' and what types of interactions invite these to help ensure clients can experience these (earlier)?
- How could ML support the identification of client struggles (i.e., false completions) of program contents?
- How could ML support the effective personalization of program contents to each person's needs? Could we give supporters an indicator how beneficial it may be to recommend/ unlock a specific module at a certain moment in time in the client treatment journey?

## 4.4 Extracting & Responding to Relevant Client Information Online and Under Time Constraints

Above, we described how supporters make use of the available information to develop an understanding of client progress or struggles; how to best personalize their treatment; and offer useful guidance. In this section, we detail on some of the challenges that PWPs encounter in gathering sufficient knowledge about their clients' situation to be able to provide effective support whilst relying predominantly on online data; remote, asynchronous communication; and working under time constraints (*i.e.*, 15 mins per online review). In particular, we highlight (i) *difficulties in extracting treatment relevant client information* where supporters either have too little information about the person; or are provided with a wealth of client data to review. We then describe *strategies* that supporters apply (or propose) for: (ii) *inviting more client input*; (iii) *soliciting key information* from client contents, and *responding in a time-efficient manner*.

### 4.4.1 Challenges for identifying treatment relevant client information in guided iCBT

A small proportion of clients (~less than 10%) can access the treatment programme without the requirement of an initial assessment call. For these so called *direct* or *instance access clients*, supporters do not have the opportunity to develop a problem descriptor statement over the phone, and may not know anything about the specific circumstances of the person (P1, P2). In the absence of much (initial) dialogue between the client and their supporter, PWPs need to extract the clients' main problem from the contents that they share in tools, messages and other platform feedback (P5). This becomes more complicated where clients do not enter much content in tools, nor leave any messages or reflective commentary in the program for their supporter (P2). In those instances, supporters often have too little information available to gain a real sense of the person (P9, P13), which makes it harder to form a 'therapeutic rapport' (P8, P9, P13) and provide personalized guidance:

> *"So they will be screened by our admin team to make sure that they're appropriate for our service in terms of their risk and whether their presentation is appropriate. When it comes into that kind of thing, I've never had the opportunity to do anything like a problem statement because I'm literally straight in, first treatment session's all written. It's very closed off but it's effective in the sense of we can improve our access to clients, if that makes sense. So there's a less rapport built because you genuinely feel like a computer software system. You know, there isn't that kind of introduction, it's all template based, if that makes sense."* (P9)

A *lesser understanding of the person introduces a number of difficulties for PWPs*:

Firstly, due to a lack of information, there is a higher risk that supporters form potentially false assumption about the person and what their difficulties might be (P5, P6). For example, P5 explains:

> *"… and obviously they don't know what they need to write to let me know that and I don't know what to ask about that unless there's something that they write that gives me that idea or that sense that's what their main problem is so that's, I guess that's happened maybe a couple of times before where actually I've made an assumption about what their main problem is but when you speak to them it completely changes."*

Outside of direct communication and dialogue with the supporter, it can be 'really hard' (P3, P4, P11) for the PWP to gather, extract and confirm that they have build-up the right information about a person. P3 explains:

> *"Depending on the client, sometimes they would write to you and say, "The last couple of weeks have been stressful for my commuting," so then you know. But sometimes then I pick up from what they put on SilverCloud, "What are* *your triggers for your anxiety and stress?" "Commute to work." So this is when I was telling you that you just have to pick that information and put them all together. That's why it's hard, because sometimes, you're not sure, "Am I getting the right message here?" If that's the case, I would then include that on my message. "Did I get that right? You left a couple of information about stress around commuting, just to make that clear, so I know if I'm supporting you in the right direction." So people who leave messages on SilverCloud, great, because we can do the job very straightforward and we know that we're aiming on the right target. But people who don't, you feel that you're being blinded of, "Hang on a minute. Is this stress from a commute? Or too much work that he's doing?""*

Secondly, without a good understanding of the client and therapeutic rapport, there is a higher risk that clients may not open up about their main mental health struggles or engagement difficulties with the program until it can be too late; meaning that the client may have either decided to drop out (P6, P12), or there was a delay in recognizing that their underlying problems (*i.e.*, bereavement or trauma) may not actually be well suited to CBT treatment (P2, P5, P7). This can have implications for both client and supporter time and resources.

Thirdly, as alluded to in the quote by P9 above, having only limited information about the client – due to instant access, or generally low engagement with the program and the supporter – means PWPs are less able to personalise their review (P5, P6, P8, P9, P13, P14, P15). As a consequence, they describe their feedback as 'more generic' (P5, P13), 'quite neutral' (P15) or 'non-committal' (P10). P14 explains:

> *"So especially just for doing online reviews, which is the majority of the work on SilverCloud, the whole point is being able to make someone's treatment individualised and make their treatment right for them, and if they're not giving you much content about what they're struggling with or what they're finding helpful, it's really difficult to personalise treatment as they go forward and to actually direct them in the right way. It means you're just very generically directing them through the programme."*

Not being able to personalise their response can leave supporters feeling that their feedback to clients is not as constructive and helpful as it could be (P2, P6, P10, P14, P15). P14 continues:

> *"I mean I have had some clients where they came straight onto the programme without having any assessments and things, and it would still work but I think again it was about how much the client is sharing with you. If you hear nothing about their difficulties… I think they can still gain out of it, but it does make it very hard for us as the supporter to give some constructive, helpful feedback and not just a very generic response about 'This is SilverCloud, work on this module now'."*

This is further bound up with feelings that one is 'not quite hitting the mark' (P5) and 'neglecting the clients' (P9), which can be one of the 'least liked' (P6) and hardest things to experience as a supporter:

> *"I think that's the hardest thing is you don't know whether you're hitting the spot because if I think about this client that I was just thinking about, the one that's having step three treatment, I don't really know why he's not engaging or messaging back. (…) you don't really know what their view is unless they're telling you." (P15)*

A lack of access to relevant client information due to little client input or dialogue also means that it can take supporters longer to write their feedback (P6), as they struggle to know what to include in their message (P11). As a result, some supporters 'review all available data' (P1, P2, P3, P5, P6, P14) 'as thoroughly as they can' (P1, P14) to try and extract relevant insights and to 'not miss out on anything' (P3). P6 explains:

> *"I think that's where it's really important to read in a bit more detail all of the other things, so even the questionnaires. So that rather than just going, "Okay, they're scoring 20 out of 27," that's where it's really helpful to look at it, because you haven't had that conversation to be able to go, "Okay. This is what you're feeling. This is what you're thinking." You're just going by if they've filled in nothing, you're pretty much just going by the questionnaires and what they've looked at. That's where the questionnaires can come in really handy."*

All this emphasizes the importance of client engagement not only with the program contents, but their supporter, in order for PWPs to be able to effectively tailor their support and provide helpful guidance. To invite **client responsiveness and dialogue with their supporter however can be more challenging to achieve online**. For one, the client may have deliberately chosen an online treatment format to avoid having to talk to people and may prefer not to share what is going on for them with others (P2, P7, P10, P15). For example, P10 explained about giving personalized feedback if there is little client input:

> *"Yes, that is really difficult. I suppose, no, and in those cases, my own, I guess, preconceptions are that you might miss out on a little bit of therapeutic rapport there if I don't know much about them. But equally, for some people, that's the absolute appeal of SilverCloud is that there's no way I could identify them. So in those cases, I guess you stick very much to the content of the programme."*

Furthermore, the communication format of remote, asynchronous messaging can make it harder for supporters to build up a connection with their clients and for them to feel the need and accountability to respond to any questions the supporter may have asked of them, or to follow their recommendations (P2, P5). P5 describes this disconnect in online messages in comparison to telephone reviews:

> *"(…) I think with the telephone reviews people often feel that they're gonna be held accountable a bit more so if they know you're calling them in two weeks and you'll be saying 'did you use this technique that we talked about last time?' they're more likely to have done that, and I guess sometimes with online reviews maybe there's a sense of they can hide away a little bit more because they're not actually speaking to someone so if they haven't done something it doesn't matter because all they're gonna get is a line in the message to say 'did you do it?'." (P5).*

For telephone reviews, contrary to online messages, supporters also frequently described that gaining a sense of the person, their feelings, and how they find the program as well as building therapeutic rapport is a lot easier to achieve (P2, P3, P4, P6, P8, P13, P14), as they can ask clients direct questions and elicit answers that give relevant insights as to what might be going on for a client. Describing how telephone reviews provide more opportunity for a client-led dialogue, P13:

> *"So, it is definitely a lot more of a conversation and a discussion if they choose telephone because with a message you can't, I'm just writing what I think rather than asking them as much. So, with the telephone it is definitely more about how they're finding it. They lead that conversation a lot more than I do. I'm there just to offer some tips and some advice really in that chat instead."*

The benefits of a more direct communication channel that enables a more personal exchange, easier bonding and targeted responses, means that some supporters described a preference for telephone over online reviews:

> *"No, it's almost policy at our place. They just say to the client "Do you want it online or do you want it on telephone, the reviews?" Personally, I suppose because I found more success in telephone reviews. It's not that I encourage it, but I suppose if I had a client which I thought might need extra support, I wouldn't encourage the online review." (P8)*

Whilst a lack of client information and supporter engagement have been reported as a predominate challenge, some PWPs also reflected on ***difficulties to review all available data*** in cases where clients show good program engagement, extensive uses of tools and frequent reports of journal entries (P3, P9, P11). For some supporters – out of fear to potentially overlook any relevant information and wanting to provide the most helpful feedback – this can mean taking extra (personal) time to work through the content. P14 describes this difficulty and their approach:

> *"Because sometimes actually when someone is extremely engaged and sending a lot of content, that can also be really difficult to manage. Because actually you're wondering are they, I sometimes find that difficult to juggle the balance of wanting to listen, liking that they're really using the programme and making the most of it, but reflecting that we have very limited time to do a review. And so, it was having to struggle that against your own feelings about how you feel about not working through everything with the client as well. (…) Yeah, I think personally I used to probably always just try go through it anyway because I was always of the opinion that if someone has really put the effort into it and it's important to try and do it, so it would usually come out of some of my own time to try to make sure to properly go through it."*

All these examples highlight the need for effective approaches for extracting relevant insights about a client. The next section describes strategies that supporters already apply, or propose, for inviting more client input as well as for soliciting key information from client contents and responding to these in a time-efficient manner.

### 4.4.2 (Proposed) strategies for inviting relevant client input and dialogue?

To provide effective guidance to clients, supporters described the importance of, and the desire for more feedback about what the client finds helpful or unhelpful about the program (P4, P9, P10, P11, P12, P13, P14). To elicit such feedback, they referred to strategies of: (i) actively encouraging clients to leave messages (P6, P8,

P10, P11, P12, P14); and (ii) asking questions to clarify any client difficulties and their impact (P2, P3, P10, P12). In addition, supporters described (iii) proposals for improving the effectiveness of client communications.

As described above, while **leaving messages for the supporter** can be rare in online clients (P3, P10, P15), having access to information whereby the client directly communicates how they are finding the programme and where they encounter struggles enables PWPs to better respond and tailor assistance to the person (P8, P10, P12, P14), which 'makes the biggest difference to being able to help them better' (P14). P10 exemplifies:

> *"I personally really, really encourage people to use the message function as well, because I know that there are prompts that they can select about how interesting they find modules, and that doesn't carry the same weight for me as a message saying, "I liked this. I didn't like this. This didn't make sense." So the message tool is really important for me to be able to tailor it."*

To ensure that clients are more likely to leave messages and that these contain therapy relevant information about the person, some PWPs proposed giving the message more structure (P4, P5, P10, P15). P4 explains:

> *"Maybe instead of the messages box, like the questionnaires, maybe there could be questions about, 'how have you been over the last two weeks?' 'What do you think of the programme so far?' 'What are your thoughts on what you've completed?' 'Do you have any questions for your supporter?' Rather than just having a general message box because then maybe it'll prompt more things out of them (…)."*

Some PWPs also suggested to include other means for explicitly deriving feedback on how useful the client finds the different program contents and how they get on with the treatment in a 'more on-going manner' (P2, P10, P11). They proposed ratings of what the client thinks of the program midway through treatment such that they can act sooner in offering something different if the current format is not working for them (P2).

Supports also expressed a desire for more insights into client's day-to-day feelings and circumstances (P1, P4, P5, P12, P14, P15) as these cannot be easily gained from the clinical questionnaires. Here, supporters described wanting to know what is generally going on in the person's life that could provide useful contextual information. Having access to more information, for instance, about the clients' social support network enables supporters to link to people outside of the treatment (P14); whilst insights into the clients live circumstance can enrich understanding of how situational factors may impact clients' mood (P1). For example, a person might be 'unexpectedly living with their in-laws or sofa-surfing' (P12); or describe changes in their employment (see Section 4.1.2 for examples). P5 describes how such insights might be achieved by asking clients to give a 'daily reflective paragraph' to learn more about how their week has been:

> *"I think again maybe just a bit of a change to the way that we, the information that we encourage them to complete regularly, for example. So it has, on the programme it has a mood diary that they can keep or an activity diary but actually maybe, we've got the journal entries but they're a bit more, they're very flexible and actually maybe just encouraging a daily reflective paragraph of what have you done today, what have been the positives and negatives of the day, or things like that, would just give us a sense of how their difficulties are impacting on them day to day and a sense of what they're living with (...)"*

Interlinked with a desire for more direct communications of therapy-relevant information, are supporters efforts to invite client input and feedback by **asking questions**, or employing motivational interviewing strategies, to expand their understanding of, and disambiguate any assumptions that they might make about: client struggles or program benefits (P3, P11; see also Section 4.4.1). Client answers to PWP questions can help clarify mental health difficulties and what might impact them (P2, P3); why there might be any barriers to, or reductions in client engagement (P10, 12); and enables better assessments of clients level of understanding of the treatment program (P10). Considering the importance of having such **direct client feedback**, some supporters proposed identifying alternative ways in which to ask clients' questions (P9, P10), with P10 going as far as to suggest making certain questions 'mandatory' – alike the completion of clinical questionnaires:

> *"The first would be – I don't know, maybe it's a bit unethical – but like a mandatory open-ended question that I could leave them, that almost presents like a questionnaire. "Can you please answer this question when you next log in for me?" (…) It means that they've definitely read my message or my feedback, or that particular, they're aware I'm asking for that."*

While client messages and feedback to supporter questions can be highly informative to solicit; PWPs also described some of the limitations of the current review process due to its pre-scheduled, asynchronous communication format. The ***fixed timing of the review can delay the receipt of relevant client information and reduce opportunities for supporters to address any problems sooner***. This has been particularly apparent in PWP reports of clients using the supporter messaging function to 'change or cancel appointments' (P4, P8, P7, P9) or to 'express a crisis situation' (P7) – possibly with little or no awareness that their supporter cannot see those messages until the day of the review appointment. As a result, one PWP described to actively discourage their clients from leaving messages to avoid such instances (P7); whilst P9 proposed developing a separate communication mechanism for clients to notify about their intent to change or cancel appointments:

> *"Say if they forget that they need to contact us directly, it'll come through a message, which isn't helpful because we won't see it until that physical day. So for them to communicate with us, we advise them to contact the service directly. Now I appreciate everyone's human, sometimes they may forget that, but say there's like a little switch button or something that suggests like, I want to cancel my appointment purely for that purpose, so we have an indication that flags up then on our [supporter interface] … Because we had the list of the clients, if it like flags up with a little like, I don't know, like a little icon or something that is purely linked with cancellation, that would then help us and save our time and save their time. But purely for that because the last thing we want is to say this client needs help and it's something really that we can deal with in the treatment session as opposed to a cancellation."* (P9)

In addition to more timely interactions, some supporters described the ***need for more flexibility in how they communicate with their clients***. For some clients, having a pre-scheduled 'supporter review' looming in response to their use and input to the treatment can create pressures – especially if they are struggling to engage. This brought forward discussions of the advantages of introducing 'instant chat' (P12, P15) as an additional functionality to create 'opportunities for having just a quick check in' in-between reviews (e.g., let PWP know *"I'm signing off because I'm feeling better, rather than waiting for a whole appointment that they just maybe are putting off or not wanting to do or…"* -- P12). P12 explains how having a more lightweight communication channel might help lower certain communication barriers:

> *"It's a bit controversial, but potentially the option for instant chat that I really see the gains and of course the challenges that come with that as well especially from certain clients that maybe struggle with boundaries and, yeah, that could be really difficult to manage. But for some people, one instant message could be a really useful tool. (…) So I've got a client for example, who was wanting to give up in therapy, she couldn't really see how it was going to work, really didn't really have any more belief in the approach was going to be beneficial for her. A lot of her difficulties are centred around the belief … so thankfully, she did engage in an appointment with me but she did say 'I didn't want to do this but I'm glad I did now because actually, I do feel a lot better on it' (...) so I think some people, especially if they're feeling overwhelmed, maybe even the idea of the whole appointment might be enough to actually put them off from engaging or attending, whereas a smaller opportunity for a chat might make them think, you know, I'm not having to invest in a whole appointment, I'm just going to let them know how I'm feeling, for example, or that I'm not really up for the appointment because then we could intervene before they potentially disengaged."*

### 4.4.3  Strategies for providing effective, timely feedback

Further to inviting client input and dialogue, PWPs also described strategies of (i) *filtering client information* and (ii) *using response templates for their feedback messages* to complete reviews in a timely manner.

To adhere to time constraints (i.e., 15 minutes to complete an online review), supporters described to quickly 'skim read' (P1 P4, P5, P6) and be 'more selective' (P2, P4) in what they focus on in their review to get 'the gist' (P1, P11, P14) of how well the client engages with the programme and what difficulties they may encounter. In ***filtering and prioritizing what of the available information to focus on***, supporters look for contents or themes that are most relevant for influencing next treatment steps (P2, P6). As described above, this includes information of relevance to the clients main presenting problem (P3, P6, P9, P12, P14) and their personal circumstances (P2, P6) such as details on family life, health, financial worries, or job changes. In addition, PWPs described looking for any words related to their 'mental health' and 'therapy' (e.g. 'I'm going through private counselling now'), which includes terms like 'alcohol', 'drugs', 'medication', 'doctor' or 'prescribed' as well as any references to 'trauma' or past 'mental health episodes'. Other information, such as changes in medication, are described as: *"really important for us to know, because sometimes medication can make you feel worse initially, so it's helpful for me to have that in mind"* (P6). To assist in their review process, some PWP suggested to perhaps consider fore fronting key client information such as their treatment history and main problem descriptor within the SilverCloud interface (P6, P12, P14) to ease access to treatment-relevant client information. Proposals for creating something alike a 'person profile' further included details that can convey a better 'sense of the person' such as: their age and demographics (P9); marital status and children (P9); interests/ likes/ hobbies (P9, P8, P14); future aspirations and meaningful past achievements or events (P8); or occupation or employment status (P9, P11). Having access to such data, whilst perhaps less treatment relevant, enables supporters to 'be more personable' in their feedback, P14:

> *"(…) yeah, so I guess in terms of the individual, anything that helps to get a little bit more of a picture of that person, whether it's, and it doesn't just have to be relevant to their treatment, it could also just be relevant to them as an individual, so we start teasing out more of those ideas that you can personalise what you're sharing with them, whether it's thinking about their network, their hobbies, things they like, things they don't like. That can just be interesting."*

The importance of access to personal client information – which we will expand on in Section 4.5 – is further apparent in the following quote by P8, who suggests clients could make themselves an avatar:

> *"So if you had some information, like maybe when they set up their thing that they could have more of a profile of who they are, what their interests are. I think maybe you could have dreams of the future or maybe aspirations or things that were significant to them in their past, like maybe that they achieved something that they were really happy about, or… (...) You know like I was saying that they should have maybe a profile, maybe that they can make themselves an avatar. That would be quite interesting."*

In addition to filtering strategies for extracting relevant client information, supporters can **make use of pre-written text templates** to avoid having to retype standard contents and save time, as well as providing them with a structure that helps ensure important information are covered. PWPs described the frequent use of five types of templates: the 'first template' describes the set-up of the SilverCloud process; the '1$^{st}$ + 2$^{nd}$ DNA' templates present as well worded standardized messages that the supporter sends if the client did not engage with the programme prior to a review; the 'good engagement' template can be used from week to week if the client is engaging with the programme, but requires adaptation (see further below); the 'treatment nearly completed' template informs the client about their last review; whilst the 'final' template is send to discharge the client.

Generally, the use of templates varies across supporters from: 'rarely' (P12) to 'sometimes' (P2), or 'often' (P8, P15); with some describing to 'only use very standardised templates – like the DNA' (P14), and to 'only use them with some level of personalisation' (P10, P11), and to save time (P4, P6, P8). P4 explains:

> *"I guess when it comes to typing the message, most of it is already there and it's just, for me, filling in my blanks, or adjusting the template itself. It fulfils me saving time and making sure that I cover everything as well, because sometimes when you're in the flow in typing you might forget to say, 'Oh, remember to complete the questionnaires next time'. That's my standard thing at the end. If they have any concerns remember to call this number and speak with me. So, I have things like that. That stays the same for every message, pretty much. Then putting in 'your next appointment is this date'. Yeah, just making sure I fill in the blanks really."*

Furthermore, template use was described as more common for trainee PWPs (P8, P10, P13; see further Section 4.6), whereas more experienced supporters often described a lesser reliance on templates, and to prefer writing their reviews mostly from scratch (P13, P14). P13 explicates:

> *"When I first started, I relied on these [the templates] quite a lot and used these probably way more than I should have done. Now I've got to know the programme a lot better and I know CBT a lot better and think it is a lot more flexible. I tend to only use these [the templates] so it fills in 'Dear so-and-so, from [supporter name]' rather than all the middle bit."* (P13)

Although the SilverCloud system provides functionality for supporters to create and store their own messages, few described having written own templates (P2, P4, P5, P6, P15). However, many supporters (re-)use specific text paragraphs (P3, P5, P6, P7, P10) that they either store on SilverCloud or in a separate document (i.e., in

word/ OneNote). These text-paragraphs for example contain additional information about topics that supporters frequently address such as: 'stress'; 'smart goal setting; 'engagement tips'; or 'additional explanations around the TFB cycle'. P3 describes their use:

> *"Depending on what the presentation is and where we're at in terms of the session, so I would have some information already saved. Say, for example if it's about stress, I'll give them a bit of psycho education around stress, or I might copy and paste that there. Then write a little bit to make it more personal to them. "So you mention that the commuter work is stressful. I can understand it, and then: stress is…" So that's the copy and paste that I've already put it together. So it's not time consuming, because they give us 15 minutes, but really, it's not enough if you think about rewriting everything."* (P3)

Despite the time saving benefits of pre-written, explanatory paragraphs that address common client problems, supporters described generally little template sharing within their service. Occasionally, standard text-paragraphs are shared on a more individual basis by supervisors within their team. Promoting the idea of a joint repository for PWPs to make use of (P5, P6, P8, P11), P6 explains:

> *"But I think if there was a way, because they are common things that come up, obviously. It won't always be that common thing, but there definitely are common things, like people look through the entire programme in one review. You go, "Steady on. Don't look through so much." Those key things I think could be helpful to share (…)."*

While templates can be essential time-savers and reduce the need to retype standardized texts, for they use, it is key that they are adapted to each client (P1, P2, P4, P6, P7, P8, P11, P13, P14, P15); otherwise they can feel as too 'rigid', 'generic', 'repetitive', or 'not person-centred enough'. P14 describes:

> *"Because I think you can read back over the message and if it doesn't feel like you've written to that person, it can just be very generic, and I think I would always think about what that would feel like for me as a client. I'd like to feel something a bit more, someone who is actually noticing what I'm doing and listening to what my problems are rather than just anybody."*

To achieve a more personalized, person-centred message, supporters described using the templates mostly as a way of giving them a 'structure' for writing the review – which parts to include at the beginning and the end – where they tend to insert the most personalised parts in the middle (P2, P4, P5, P6, P8, P13). P5 states:

> *"So it might follow a similar pattern in terms of those standard templates, but yeah I would always bulk it out in the middle so I might start by saying which is in line with the template, 'I hope you're well, well done for logging in three times since our last review, I can see that you've now completed X and X module, I hope you found these useful' and then that's when I would go into my more tailored stuff, and then at the end it says, 'the next module I'd encourage you to work through is … and that will cover this… and I've set a new review date' etc so the bare bones of the templates I still use, but I always would fill it out in the middle with those kind of more personal responses and things like that."*

The next theme will detail on the strategies that supporters employ to provide a more personal response.

### 4.4.4 Summary & Opportunities for ML

In summary, this section discussed some of the challenges that supporters encounter in assisting their clients, especially in cases where they do not have much information about the person to provide effective, personalized guidance. A lack of easy access to most relevant information about the client carries higher risks that (i) PWPs can form potentially false assumption about them and their difficulties; (ii) therapeutic rapport can be harder to establish, which can reduce opportunities for clients to open up about their mental health struggles or engagement difficulties; (iii) PWPs are less able to personalise their review and provide constructive, helpful feedback; or (iv) where too much data is available to review, that it takes supporters too much time to filter through client content. All this highlighted the need for effective approaches for engaging clients in dialogue with their supporter and for accessing and extracting relevant insights about them.

We described, how such exchanges however can be more challenging to achieve through remote, asynchronous online communications; and outlined some of the strategies that PWPs already employ to solicit direct client feedback (i.e., by inviting online messages and asking questions); as well as proposals to increase the flexibility and effectiveness of current online communications. Interlinked with this is the need for better access to relevant client information and to provide feedback in a *time-effective* manner. To this end, PWPs explained their approaches to prioritizing client information that are most relevant for influencing next treatment steps and to filter client contents accordingly by looking out for important keywords and details about the person and their life circumstances. To save time, supporters make use of pre-written text paragraphs and message templates, which are often written by themselves, but rarely shared amongst colleagues or services. The message templates are mostly used for very standardised responses (i.e. DNA events) and where PWPs are less experienced in giving feedback. They serve to give structure to the message, but require personalization to not be perceived as ridged or repetitive; and to achieve the all-important person-centred feel.

Responding to these tasks and challenges, ML research might explore the following questions and opportunities:

- How could ML help extract/ foreground relevant client information (i.e. language indicators related to medication, treatment episodes, etc.)? Could we auto-populate a person profile over time?
- How could ML provide additional feedback to supporters to aid their understanding of what the clients main problem is and what about the treatment is or isn't (potentially) working for them?
- How could ML invite more opportunities for client input/ feedback and dialogue? For this type of client, what communication strategies would make it more likely for them to respond to their supporter?
- How could ML reduce time-consuming labour (e.g. auto-complete generic summaries) for example as part of template uses? How to translate individual templates as a machine-readable resource?

## 4.5   Forming a Therapeutic Alliance for Improved Therapy Outcomes

In this section, we describe (i) the *importance of 'personalized support' and establishing a therapeutic relationship between PWPs and their clients* for improving client engagement, disclosure and treatment outcomes. We then report (ii) on the *strategies that PWPs described for providing personalized support*.

### 4.5.1   Importance of 'personalized responses' & 'sense of the person' for establishing a therapeutic alliance

When we asked supporters about the relevance of providing personalized feedback to their clients, they emphasized its general importance as well as the need to gain a 'sense of the person' for enabling the *formation of a therapeutic alliance* (P1, P2, P4, P5, P6, P7, P8, P9, P12).

Personalised responses to the client were described as essential for conveying that the supporter is a 'real person' (P1, P2, P7, P12) and a 'real therapist' (P5) with 'accredited skills' (P10) who takes the support role 'seriously' (P2). Tailored feedback and expressions of empathy and 'understanding of their situation' (P12) further demonstrate that the PWP is 'listening' and 'interested' in the client (P1, P2, P3, P5, P6, P7, P11); which are ***important factors for the client 'to feel supported' (P6, P8), 'to feel part of the treatment' (P11), and 'cared for'*** (P1, P2, P6, P12). P1 explains the role of personalized feedback especially within online treatment:

> *"I think it's just to make sure that the person feels heard and the person can feel that even if using an online platform, that this is…, that we're still working in a caring way to support that patient and we ultimately want them to get better, too. So, I suppose showing that care means we're making that human connection and we want them to get better and we want them to be supported. I suppose that's why it's so important."*

Simultaneously, conveying a 'sense of human-ness' and 'human connection' through personalised responses has been described to help ***reduce perceptions of the supporter being perceived by their clients as an 'AI', 'robot', 'auto-responder', 'computer or mechanical system'*** (P1, P3, P4, P6, P7, P8, P12). P12 states:

> *"So for me, again it's about the client feeling it isn't just an artificial intelligence and someone who's not human responding. I want them to know I am a human and I have got empathy and I can relate, and I can understand their difficulties and reassure them. So if you don't make that personal, then you know, they could put something in there like, they've had a difficult day at work and they ended up going sick the next day because of it, and if that's not acknowledged, I think for me, if it was the other way round, I think are they even listening to me? Do they even understand the kind of week I've had? So for me, picking up on those things is really, really important, yeah. (…) Because again, from my understanding and learning, the core part of building a therapeutic relationship which is supposed to enhance someone's experience and journey through therapy and build confidence in your intervention* *is by showing that compassion and care and building that empathy and understanding of the client. So for me, I think the lacking of that might undermine that process from happening."*

Having a connection and rapport with another 'human' as part of the treatment process was further interpreted to **instil 'hope and trust' in clients that the online intervention can be helpful** for them (P9), and reduce any reservations that they might have about the use of an online therapy approach (P1, P3, P7). P1 describes:

*"You want them to feel like you're reaching out to them as a person, even though it's an online platform. And I think that's why sometimes people do have reservations about using an online programme. They're like 'Oh, I want that human interaction'. That's important to the user, so it's really important."*

Thus, especially for online clients, who do not receive telephone support, PWPs emphasized the importance of a person-centred approach and to add more personalization (P5, P7, P9) so as to **ensure that their clients receive the 'same level of support' as they would in other forms of treatment** (P3, P5, P13):

*"And I think if it was the other way around, I would want it to be personalised. I wouldn't want a therapist just spending two seconds sending me a template. I'd want to feel like it was coming from their knowledge of me and what's going on for me, just the same as it would be a telephone. It should be equivalent. (…) I guess because we're offering both as equal forms of treatment. I guess if we were expecting different results or different, …my engagement to be different, then I would be a bit more choosey as to what clients I put where. But because we offer it up to the client as their choice, I don't feel that they should get a weaker version of me just because they've got a full-time job, they've got three kids and they're very, very busy, but they've had to opt for online to engage and they get less of me than if they had a bit more time and could have a telephone call." (P13)*

Moreover, PWPs suggested that a therapeutic alliance – nurtured through personalized messages that reflect their listening, support and care for the client – is crucial for **maintaining and motivating client engagement** (P2, P4, P5, P13). P2 shares: *"If people don't feel that you care, if people don't think someone's there to actually help them, then they're not going to engage."* Here, engagement includes program use and also clients' responsiveness and readiness to open-up to their supporter (P1, P4, P8, P9). P9 states:

*"But at the same time, if you are direct with them by using their kind of language or if you're, as I say, tapping into those personal traits or characteristics, they come back with you, come back with a thanks for your help, I've had a really tough week, they're more open and honest."*

As a result, a good therapeutic relationship has been explicitly described by PWPs to help **improve the treatment process and overall therapy outcomes for the client** (P3, P4, P5, P7, P13). P5 summarizes:

*"Because we know that a good therapeutic rapport is conducive to better treatment outcomes. That's obviously one of the main reasons and that you just want them to have a positive experience as well. You want to have a good experience with the programme, you want them to have a good experience with the service, and it does I think help people to engage better, it does give them that sense of hope that what we're doing works. It can make them feel listened to. For some people just feeling listened to is quite powerful in itself to help them so, yeah, it's really important and it has, there's many reasons why it's important and we know that it does just help better outcomes, we get better recovery rates and things like that with it."*

Importantly, to establish a therapeutic alliance, **supporters need to have a 'sense of the person'** (P6, P8) – a snapshot of what is going on for them; who they are; how they've been and are getting on with the program (P6) – such that the PWP can be more personal and invested in their response. Without a 'sense of the person' it is harder for supporters to be able to relate to the person and provide effective, personalized guidance (see Section 4.4.1 above).

### 4.5.2 Strategies for providing patient-centric "human" care

In addition to the personalization strategies described above that depict how supporters provide feedback to clients to help overcome program obstacles (Section 4.3.2) and that seek to motivate, reassure and encourage client engagement (Section 4.2.3); PWPs described **their explicit efforts in online communications to convey a sense of them being a 'real person'** and **ensure the client 'feels heard, understood and cared for'**.

To convey this, PWPs suggested a number of different strategies. This includes: (i) *referring to clients and key people they mentioned by name* (P9), as well as (ii) *using their own name when signing off on their messages* *and having a profile picture* associated with them (P14, P15). They also mentioned that: (iii) *responding at a specific time rather than immediately* would reduce perceptions of them being a robot (P15); (iv) using an *informal writing style that reflects how their would speak* to create a more friendly feel (P4, P14); and how (v) *mirroring clients language* such as clients' use of smileys or specific words in describing their experience to aid perceptions of the PWP being 'more personable' and to promote client understanding (P2, P5, P9). P4 explicates:

> "I would write how I would talk in the call. Maybe it's a bit more informal but if I was to put myself in their shoes, I think I'd feel a bit more comfortable receiving a message that sounded a bit more friendlier rather than formal. (…) Whereas if I typed it how I was speaking maybe they'd be more likely to respond to it. That's how I like to think anyway. And I have had good responses from that in the past so I've just taken on board that that maybe works. I've gone ahead with it."

In writing their responses to clients, supporters also described the importance of 'leaving personal rather than generic feedback' (P1, P2, P8). To achieve this, they described strategies such as (vi) *asking the client personally relevant questions* (P1, P6, P7, P14) that may relate to things that the client mentioned in a previous sessions (i.e., an upcoming birthday), messages or tools (i.e., difficulties with family or work). P7 states:

> "So I would ask…I usually put in the review message 'What have you been up to in the last couple of weeks? How have things been?' If work was a difficulty, I'd ask them about work or if it's their daughter that was sick last time I'd ask them how their daughter was. Just making it a bit more personal really."

The vast majority of PWPs (P1, P2, P3, P4, P5, P6, P7, P8, P9, P11, P13, P14, P15) described additional strategies of (vii) *reflecting and referring back to the client what they had said or done previously* to help create more of an emotional attachment (P9). Perhaps in its simplest form, this involves repeating something that the client had written (i.e., in their journal or tools) to indicate their entries are read (P2, P13). Or, PWPs may refer to things the person mentioned during an initial assessment call (i.e., personal interests mentioned); or comments with regards to their personal life circumstances (i.e., things they encountered that they found hard). P1 explains:

> "If we're doing a depression module and it's about looking at activity levels, 'You said you used to really enjoy cooking. Have you had a chance to look at that again now that you've had a chance to read this?' because then they remember, they're like 'Oh, that person really listened to me at the assessment'. So, referring back to previous information can be really helpful, too."

Making such explicit efforts to show how they are *listening and remembering specific information about the person over time* by relating to previous conversations (P3, P4, P8, P9) helps demonstrate continued interest in the clients' lives and are key for enabling the formation of a personal connection. P3 describes:

> "I remember when we first spoke over the phone, you mentioned about this. I wonder how it is now. You were worrying about that. Are you still worrying about this?" So having that element to connect on the previous conversation or messages is always helpful. Like I said, that's how they feel that you're connected to them (…)"

Moreover, to really make the person feel heard, understood and cared for – which is really important in therapy – PWPs described strategies of (viii) *recognizing and normalizing client struggles* (P4, P6, P8, P11, P12, P14) as well as *responding with empathy and compassion* (P1, P5, P6, P14). For example if the client mentioned something upsetting, the supporter would engage with that and 'not just brush over it' (P14); or they would acknowledge specific client struggles such as lacking 'the will power needed to stay motivated' (P8), 'fitting the program in one's life' (P4), 'reading a lot of materials' (P11), or experiencing a 'difficult patch' (P8). P8 illustrates this approach as follows:

> "Or, if they're going through maybe a difficult patch and things have got worse again then I would give them some kind of support around that saying things like, "I know this is really difficult and maybe that the road to recovery is not always a smooth road, and it's especially motivation, it's a bit like pushing a car, it's hard and it takes a lot of effort and actually down the road that you might have a few bumps so it's not going to be smooth, but actually once you get that momentum going then it will get easier." So, I might say things like that."

P6 further explains the benefits of normalizing client struggles and responding with empathy and compassion for conveying that sense of care and personal connection to their clients:

*"I think it is. I guess that's where the normalising and things like that come in. It's that sense of you're really caring for this person, you are connected to them. I think otherwise it can just feel very robotic. For a lot of our clients, they can be relatively hesitant to go online, so I think those key things make them see that actually, this is good form of working and that we do still care about you, it is still personal."*

### 4.5.3 Summary & Opportunities for ML

In summary, this section described the role and importance of personalized support in establishing a nurturing, person-centred relationship between supporter and client. The resulting therapeutic alliance offers key benefits, including: (i) reduced perceptions of support being delivered through a machine rather than a human being who listens, understands, and cares for the person; (ii) improved client trust and hope in the effectiveness of an 'online' approach; (iii) assurance that the same standard and quality of care is delivered via an online treatment as is in other therapy formats; as well as (iv) better client engagement with the program; and (v) responsiveness to their supporter, which they can tend to mirror. All these factors promote overall therapy outcomes. To establish a therapeutic alliance, PWP described providing personalized assistance through explicit efforts to convey in their communication a sense of them being a real person and ensure the client feels heard, understood and cared for. Amongst the strategies they applied are: the use of person-identifiers such as their own and other peoples' names; the timing and writing style of their messages; asking the client personally relevant questions; reflecting and referring back to them what they had said or done previously; as well as recognizing and normalizing client struggles and to respond with empathy and compassion.

How could ML enable supporters to get a better 'sense of the person' (to care & connect)?

- How could ML advance supporters' understanding of what types of their communication strategies (i.e., encouragements, normalization, etc.) are most effective for a particular client?
- How personal/ personable (or frequent) should PWP feedback be to convey a sense of 'human care'?

## 4.6  Upskilling Supporters & Boosting their Professional Confidence

In this section, we describe (i) some of the *challenges that novice supporters encounter in getting started with SilverCloud* as a digital therapy offering, which can prolong the time it takes for them to build-up the skills and professional confidence to be effective in their role. Corresponding to these challenges, we outline *key competencies and configurations of supporter feedback* that PWPs characterize as important in creating effective, patient-centric messages that can inform future learning and training efforts. We close by highlighting (ii) the *importance of PWPs receiving feedback about the successes of their practices for boosting their confidence* in decision making around client care, and for their motivation and enjoyment of their work.

### 4.6.1  Key challenges & competencies for skilfully guiding clients through SilverCloud

As shown in Table 1, the vast majority of PWPs who we interviewed described their supporter expertise as "intermediate/ expert" or "expert". In the interviews, they described that it took a fairly long time, mostly 6-12 months or more, for them to reach this experience level and feel confident in supporting their clients with SilverCloud (P1, P2, P5, P6, P7, P8, P10, P11, P12, P13).

As factors that helped speed up the development of their skills, they described the presence of '***existing CBT or counselling skills***' (P1, P2, P4, P9, P10); and access to a '***good mentor***' (P3). Describing their journey, P1:

*"I think realistically probably it took me about six months. But I don't think that was down to SilverCloud; I think it was down to me being new as a trainee in my role in its entirety, does that make sense? (…) in my previous [role] it was just completely new learning CBT full stop. So, I don't think it's a reflection on SilverCloud necessarily, I think it was down to my general skill level. I think the more understanding I have of CBT interventions, the easier it is for me to pick up. So, I'm definitely more confident now using SilverCloud even in my new role than I was using SilverCloud in my previous role when I first started."* (P1)

Furthermore, to get more practice sooner, PWPs emphasized the importance for supporters to '***prioritize work with online clients***' on SilverCloud (P6, P7, P12). However, especially novice supporters can often be 'reluctant, hesitant and apprehensive' (P6, P8) to use online reviews. They can find it difficult to know how to use their therapeutic skills effectively and build-up rapport through online communications (P6), and thus, choose to conduct telephone reviews instead. P8 describes the lower priority of online reviews:

> *"I think that the problem is everybody expects them to be hitting the ground and running. I think that that's the problem, is I think that because it's online and you're not seeing the clients then it goes down the pecking order of your priorities of learning and adapting and so therefore the skillset isn't built that quickly because you have too many other priorities."*

Amongst key learning challenges and competencies of a SilverCloud supporter is the need for them to '**be very familiar with the programs**' on offer (P1, P2, P3, P4, P7, P9, P10, P11, P13, P14). Initially, novice supporters can feel overwhelmed by the breadth of content and tools of the various SilverCloud programs (P12) and additional materials such as templates and program manuals. Familiarizing oneself with all this information, not only from a support point-of-view, but also from a client-perspective (how they come to see and experience the service) whilst keeping up-to-date with any changes or program updates, takes time (P3).

> *"Even though we have our accounts to log in and look around on how to use the system, I know we can invite ourselves so we can see what it's like on the patient side of things. But, it's having the time to do that and actually seeing what it's like for them."* (P4)

Yet, familiarity with the program contents and tools is key to effectively make use of the resources available, to not be tripped off by the odd question asked, and to provide the best possible support (P4, P7). P3 exemplifies:

> *"(…), at the very beginning, [I] was not taking the opportunity of what SilverCloud can offer me. So often, I would forget that, "Oh, I can unlock that module." Instead, I would then attach separate information when it's already there. So I guess just having your own account helps, because then you can see what's in there as well and giving your time. I didn't give myself some time to go through the modules and see what's in there. I've just learnt it over time. So if you allocate the time for that and make sure that you do it, then it saves you a lot of the time that has been wasted with me during the process."*

In addition to building up familiarity with the programs, it takes time to **develop experience** in explaining the contents (P7), and to be confident in knowing the answers to any client question (P13). Through repeat experiences of how clients typically respond to their feedback, PWPs develop their understanding of what tends to be helpful (P4, P8) and gain confidence in their practice. P7 explains:

> *"I think it's kind of how I'd…it's easier, I guess my confidence is something that was better, how I explained things, if anyone was coming up with any technical issues I'd be like 'Well, there we go', or if they were lost. Because I have my…like a demo account, so I went through that myself, so when people said 'Well where is this [name of P7]?' I'd go 'Well, there we go.' I don't have to go back to my supervisor like maybe some of my colleagues will or ask a colleague, I could just go 'Well it's up on there or it's here', so there's that kind of confidence in myself and knowing what I'm doing. That helped I suppose."*

Supporters described spending most of their time on either the 'writing' (P4, P9, P14, P15) or 'reviewing and writing of their personalised review' (P2, P5, P7, P11, P12, P13). While generally described as good use of their time, learning and **knowing what to write** in their messages and **how to be personal in the feedback** whilst adapting to different kinds of people was described as most difficult (P2, P4, P10, P11). As a result, novice supporters admitted a tendency to rely on templates a lot more in initial stages, and to not personalize their feedback as much (P12). About the challenge of not knowing what to write, P10:

> *"Just knowing what to put in the messages, I think. Just knowing what's the best thing to put in the messages, what kind of feedback I should be giving."*

To achieve a personal response, more experienced PWPs emphasized the importance to '**read client content carefully** to avoid misinterpretation' (P7, P14, P15); to always look at client tool entries (P7) as well as review their engagement, content views and clinical scores (P1). P1 explains how they pick up on relevant client cues:

> *"It's just making sure that you are personal when you review, that's a really big one, and just making sure that you are picking up on the cues that patients leave or don't leave. That's the trick, really, with an online intervention. So just making sure you notice any changes in activity and making sure that you've read the comments thoroughly enough for anything that patient is really struggling with or not. They're the real tricks to using SilverCloud."*

Further, PWPs need to learn how they can best *make use of their already existing common factors skills*: that is: identify ways in which to express empathy, praise or strategies like normalizing client behavior through written online communications rather than with body language or vocal tone (P6, P12, P15). P6 states:

> "A lot of the common factor skills we think of are those things for face to face and telephone. Like non-verbal cues would be saying things like, "Yes," and saying things like that as you go. They're thinking, "I can't do that online." Or good eye contact, good body language. It's just like you're removing some more of those skills that you could use, because you're going online, which I think makes them [novice supporters] feel… A lot of them come in and they're already feeling a bit anxious about telephone working, because they expected to be face-to-face. They think, "They won't know what I'm doing, so how can I convey that on telephone?" I think because you're taking away that, "Oh, yes, that must sound really difficult," and that nice empathetic tone in your voice, sometimes people struggle to know how to translate that to online."

It was further emphasized that expressions of a sense of care can be achieved by *acknowledging the person* and *what they've done on the program* (i.e., summarize their progress – P1), and therefore do not require for the feedback message to be very lengthy (P10) or include many empathy statements (P10). In contrast, PWPs described the importance to not get hung up on loads and loads of comments as this can feel overwhelming to the client; and instead to pick up and *prioritize 'the most important issues'* and things that motivate the person, reminding them that the supporter is there to help (P1, P5, P6, P10, P11, P12, P15). P5 explains:

> "I know that when I support new starters and trainees, they do have a tendency to try and respond to everything that's on there or everything the client's written and actually that's not always entirely helpful so I, yeah, really try and… I have a look at sort of everything but pick out the bits that are relevant or the bits that I feel kind of need a response, that thing."

Considering the importance of continued client engagement, supporters also highlighted as a good practice to *actively invite client feedback and dialogue* by asking the person questions (P6, P13), letting the client tell the supporter what they find helpful and how they want to be supported (P9, P10), as well as to not be scared to reach out to clients who don't engage (P2). Further emphasized was the importance to **be positive and encouraging** (P10, P13), even if clients did only very little on the program. P13 expressed:

> "(…) and that encouragement, I guess, a little bit. That even if they've only done a little bit rather than focusing on 'You haven't done very much' thinking 'Oh, great, you've done this. Let's focus on this and we can build on it for next time'. So that more positive viewpoint on it rather than a negative slant."

Finally, supporters should be reminded of (i) the benefits to *carefully manage client expectations* from the outset, such that they know that much of the learning and practices is expected to be done by them through self-work, and to pre-empt any false expectations of any continuous communications with their PWP outside of scheduled reviews (P10, P13); (ii) the need to *keep an eye on careful time management* by avoiding to retype repeat content and using templates where feasible (P3, P6, P15); as well as (iii) their role to *provide clients structure and a direction* forward through the treatment program (P10).

### 4.6.2 Boosting supporter confidence: Desire for more & frequent feedback on practices

When we asked supporters what they **least liked about their work**, they described as main difficulties previously mentioned struggles of 'user disengagement' (Section 4.2.1), client 'misfit to the treatment' offered (Section 4.1.1), and having 'too little context information about a client' for providing effective, personalized feedback (Section 4.4.2). In contrast, as **most enjoyable about their work**, they described being able to see and receive feedback that their clients are 'really engaging with SilverCloud' (P5, P14); that things are 'really changing and improving for them' (P1, P2, P5, P6, P7); that they are having a 'really good patient experience' (P7); and that the support offered to them was 'really helpful' (P3, P5). P7 explicates how this can provide them with a sense of satisfaction, validation and assurance in their practices:

> "I guess their feedback, I guess their feedback kind of helps with satisfaction and kind of getting that validation from what they are leaving or the messages or seeing how they are finding it helps. I suppose that's something that feeds into that validation or assurance or whatever you like to call it."

Simultaneously, however, PWPs also expressed often feeling uncertain about how their support impacts their clients and whether their actions truly benefit the person (P7, P10, P11).

> *"I think sometimes the rewards aren't as tangible if you don't know the person and you can't visualise them. Equally, if they haven't left you any messages to say, "Thanks. That was really helpful. I'm really glad you mentioned that," then, yes. You can question what you contributed to that, for sure."*

One of the only ways in which PWPs get explicit feedback about their practice is through the *patient experience questionnaire* that is sent to their clients at the end of treatment (P7). Yet, this offers little insights on how clients get on with the CBT program during treatment. Here, supporters expressed the desired for 'earlier' (P2) and more on-going feedback whether their messages are being helpful or not (P2, P10, P11); as well as more information on the factors that may have led, for example, to a client dropping out from treatment (P12). P2 describes the importance of feedback on their support practices as follows:

> *"I think it would be helpful, again in terms of supporting people, I think it would be helpful to help them engage and to enjoy it kind of quicker but I suppose in terms of what we do, yeah, it would be nice [to see more feedback from clients]. If people have kind of good experiences or good things to share, we always want to hear that because obviously a lot of what we hear unfortunately, is doom and gloom and the worst situations for people. So good things is always nice to hear. Not me personally, I'm not so fussed about hearing whether I am being a good supporter in the sense that oh, yeah, you've been fantastic, and we've really enjoyed it. I'd much rather hear this has been helpful or the way that these reviews were done were helpful or actually, next time people might prefer it this way and I suppose the good points of the job and kind of the constructive criticism as opposed to you personally have been brilliant, but yeah, no, good feedback's always nice to hear."*

This raises the question how applications of ML might be able to contribute to PWP's desire for more, and more frequent, feedback on their practices; and how having access to information as to how their actions can benefit their clients might help build-up their confidence.

### 4.6.3 Summary & Opportunities for ML

In summary, in this section we brought attention to some of the challenges that especially *novice* supporters can encounter when getting started with Silvercloud, highlighting how it can often take a fairly long time for them to develop their support skills and confidence with this treatment format. Barriers include the need for familiarisation with the details of expansive and changing treatment programs; hesitation and inexperience with using an online approach that requires PWPs to translate common factors skills into written communications; and knowing how to effectively personalize feedback messages to each person. To address in particular the latter barrier, PWPs highlighted as good supporter practices to: carefully read client content; be positive and encouraging; acknowledge the person and their progress; invite client feedback and dialogue; focus on only on the 'most important issues'; and to provide clients a clear structure and direction through the program. To better manage the review process, supporters also need to carefully manage their time, and often benefit from setting clear client expectations of their respective roles early in the treatment process. Lastly, supporters described the benefits and a desire for more feedback about how their actions impact clients, which can, i.e., clarify reasons for client drop-out, or help improve their confidence in decisions around client care. This can enhance treatment effectiveness and add to PWP motivation and enjoyment of their work.

Responding to these learnings, ML research might explore the following questions and opportunities:

- How to assist especially novice supporters to communicate most effectively and build-up their confidence in feedback message writing (e.g., by providing smart guidance on "what to say" when/ tailoring communication style)?
- How to help supporters better understand how their actions matter: what types of actions (and their timing within the care pathway) achieve positive change/ have the most impact?

# 5    Design Sensitivities for Assisting PWP Practices (through ML)

In this section, we describe how discussions of initial ideas for new or additional data representations about client behaviors or outcomes started to highlight important sensitivities for the design of machine-generated insights as part of supporter-led client review practices:

*Efficiency of Process & Adding Value through Data.* Many of the client behaviors that supporters should look out for, are already summarized at a glance via usage metrics – including clinical scores, number of logins, pages viewed, and client notes. Other information is more complicated to assess: has the content the user reviewed been relevant to them and their situation? Have they completed the program exercises, been reflective and made progress? Do their responses indicate any patterns in thoughts or behaviors? And to what extend are clients' confident in using the program, and know what to do next? This raises the question as to how alternative data representations and uses of ML could facilitate such assessments. In expanding the amount of data and insight offered, we need to be mindful however to not overburden supporters simply by providing 'more' information. Being limited in time to review each client requires *efficiency* in information processing; thus, the information needs to enable supporters to provide their feedback faster, or to create *additional value* for clients (i.e. by achieving a more personal message).

*Empowering, not Replacing, Supporters in their Practices.* Assisting supporters in their practices however does not solely present an 'information access' challenge. Beyond, we need to consider how the inclusion of (ML-) data insights and related (system) recommendations may shape supporters understanding of their own role and contributions to assisting clients as part of a computerized intervention. Our research highlights how supporters really enjoy, i.e., the 'unlocking' new modules for their clients as it gives them a sense of agency and purpose in their desire to help people. If, however, based on advance data processing, the intervention was set up to automatically propose an 'optimal' module to recommend to a specific client, this could be perceived as reducing supporters' input and value in favor of technology overcoming human information processing limitations. Thus, to truly empower supporters through novel data insights, it is important to identify ways in which ML-enabled insight can assist them to provide their services more effectively, and build-up their professional competence, rather than trying to replace or compete with them. This requires careful exploration and study of how (ML-) data provision can form part of their work practices such that it can help increase the scope of productive actions that they could take, or help them feel more confident in their recommendations to clients. Interlinked with this are the following sensitivities:

*Not Losing Human Touch & Fostering a Connection*. When thinking about ML analysis and ways in which to optimize work practices, one might quickly review the current use of templates by supporters as an obvious route for standardizing communications to make these more effective. However, such efforts might also risk (i) reducing the personal look and more handcraft-y feel of personalized supporter messages; and (ii) conflict with a preference for the openness that is currently provided to supporters to bring in their own patterns and communication styles. The nature of remote, asynchronous messages, facilitated by template structures, has led some clients to question, in previous user feedback reports, the authenticity of the supporters' identity as real humans. Thus, given the described importance of nurturing a therapeutic alliance, this raises the question on how best to leverage the presence of a 'human-in-the-loop' and increase clients' awareness of the supporter as someone who is invested in them and their therapeutic success so as to foster that relationship and associated benefits. How might ML approaches best support and build on supporters' current practices in connecting with the person and tailoring their messages?

*Feedback on Efficacy of Support.* In supporting SilverCLoud clients, the role of PWPs is not to provide therapy, but to help guide the person in using the platform, so that they learn how to make themselves better. As a result, supporters are less involved in the actual learning process, and they lack many of the natural feedback channels that are available in other treatment formats such as face-to-face therapy. This, in combination with having less information available by online clients that show little engagement and content sharing, can make it difficult for supporters to understand if and how their messages are beneficial to their clients. Often motivated to want to make a positive difference in peoples' lives, there is thus a need to make more apparent how the help offered has or might impact their clients so as to improve support effectiveness as well as PWP confidence.

# APPENDIX 1: Advantages of Guided iCBT Support with SilverCloud

When asked what supporters liked most about using SilverCloud to support their clients, they described it as 'easy-to-use' (P2, P3, P6, P8, P9) and valued the interactive nature (P3, P6, P7, P12) and aesthetic appeal (P9) of how psychotherapy content is presented. In particular, they described the qualities of the platform as a 'knowledge base'(P1). This included an appreciation for the 'depth, breadth and variety' of the program contents (P10) and the 'flexibility' (P3, P4, P7) provided through the availability of 'different interventions and techniques' (P5) that allow supporters to personalise treatment to people with 'more specialized difficulties' (P14). For example, P2 described about the ability to choose amongst different treatment programmes:

> "(…) if someone's got social anxiety, right let's change the programme. I can make that specific for you. If someone's got diabetes, great, I can make that specific for you and I think that again has been helpful in helping me to engage people because I have the better resources to do that."

As a knowledge base, the various programmes offered through SilverCloud were described as 'very informative' and to provide more information than what a supporter could go through with a client, especially when compared to other treatments like 'guided self-help' (P14). As such, the supporter does not need to focus on explaining the interventions, but can provide additional knowledge and skills to help contextualize or expand on what is already there on the programme (P3, P4, P5). This frees them up to give further support to their clients (P13, P15) and save time (P5, P11) since a lot of the actually learning can happen outside the supporter review (P11, P12). At the same time, supporters can make use of so called module sheets that provide them with context information on a certain client presentation, or explain a specific technique, which can help guide them in their practice (P3, P12), especially when working with clients whose presentation occurs less frequently or is more complex. P5 for example explains:

> "A lot of your work is done for you in terms of the explanation of the interventions and the rationale for using them, so actually yeah a lot of the nitty-gritty of the interventions is on there for them to read and then we're just there to top up that information so it is less time-consuming and makes our job a bit easier."

Furthermore, in having to incrementally build up their own knowledge base and practice self-learned skills in-between scheduled supporter reviews (P1, P13), clients are given more 'independence to take responsibility of their treatment' (P1), which puts 'a bit more onus on the patient' (P1, P12, P13). P12 describes this as follows:

> "It's a bit of a shared ownership of the therapy role, which means, I believe it gives the client a bit more confidence in their abilities to be more independent with their future gains from therapy …and not become dependent on a person and you know, yeah, can I do it myself because I already have almost".

This sense of a shared ownership is also reflected in supporter's descriptions of their own role. They characterize their work as 'collaborative' (P10, P13) in aiding clients' therapy (P14), and their role as someone who 'guides and facilitates' client progress through the program (P15) rather than a 'teacher' or someone who would 'directly tell the client what they need to do' (P13). They achieve this by engaging the person in conversation to provide a 'personal side' and 'motivation' (P13), 'supporting and empowering them to make changes that improve their wellbeing and lives' (P5). In their own words, P10 describes:

> "(…) this is a self-directed learning platform. So my role is to support them and help them see the relevance and tailor and apply the programme to them. If they're not then going to do what I recommend, whether that's recommended through an online message or a telephone message, there's not really a lot more I can do. I work with adults, and there has to be a point at which this has to be very collaborative. We do what we can and we tailor it as much as we can and we're as engaging and as open to conversations if they're struggling as we possibly can be. But if they don't try it and they don't do the work, that's up to them."

What further facilitates supporters' work practices is the fact that everything the client has done with the programme is all 'in one place' (P2, P3, P8) and visible to the supporter (P6, P10). This adds to 'consistency of information' (P3) sharing between client and supporter and means that supporters do not rely on clients' word and telling how they engaged with the intervention materials (P7, P12, P13, P14). As an important information resource, supporters also found it re-assuring that clients can keep access to the service for one year post-treatment (P3, P6, P7) through which they can keep accessing the information they have learnt and carry on using the techniques and feedback offered.

Finally, adding to a sense of client empowerment, supporters further described the ability for clients to track their own progress and getting direct feedback (P10); that they can work through the program at their own pace and time, and fit it flexibly around their lives (e.g., especially for mums, shift workers etc. who may find committing to a specific appointment challenging) (P6, P15); and that clients have the time to think through their own examples and difficulties before a supporter review, rather than having to do so on the spot as can be the case for other forms of treatment (P7).

In summary, supporters particularly valued that SilverCloud provides a rich knowledge base of therapy contents and techniques that can be flexibly tailored and that facilitate self-guided learning by clients outside of supporter interactions. This contributes to a sense of shared ownership of the therapy role, whereby clients take responsibility for their own learning and progress, whilst supporters provide them with a personal connection to help guide, motivate and empower them to make the changes that will improve their mental health. Having all program information available in one place and shared between supporter and clients further contributes to transparency about client engagement and progress, and promotes consistency and flexibility in information access.

# APPENDIX 2: Initial PWP Feedback on Outcome Prediction

In conversation, four supporters (P1, P2, P8, P14) were asked about how they would feel about having an automatic prediction of their clients clinical outcome trajectory given to them. Their feedback was mixed. Three saw value in having a prediction that someone may not engage and benefit from the treatment in that it would enable an **earlier identification of client problems** (P2, P8, P14), through which other, more supportive options could be offered to the client sooner, which will avoid wasting time, supporter resources, and for the client to not get the help they need (P2). P14 describe most use of such a functionality to pick-up client difficulties with the program earlier; ideally already during the assessment stage:

> *"Because I think the main thing is that the people, the quicker that someone isn't getting it or isn't engaging with it, the quicker you're picking that up the better, because the quicker you're moving them on to something that's going to be more helpful to them. And I think it's always a case that the longer sessions you go without getting, catching the problem, the worse it's getting."*

The same supporter can also see how outcome prediction could be helpful as an information resource during case supervision by providing the supporter an earlier indicator of client struggles.

Despite these positives, the other two supporters also had **reservations about the usefulness** of such an indicator. P1 describes concerns and disbelief about how realistic such a prediction could be; and both P1 and P8 described worries of how having such a prediction could set-up specific expectations and preconceived ideas about the client in supporters that may be unhelpful (e.g. reduced supporter effort knowing that client is unlikely to benefit). P1 states:

> *"I suppose for me personally I don't really believe in that because it's almost like, you know, it's forecasting in a way that I think puts almost a false, it's a calculated expectation of someone and actually how they… I don't believe that how they will progress or won't progress can realistically be predicted. So, for me I don't think that would be necessarily helpful. That might just put our own preconceived ideas on that person, 'Oh, they should be progressing' you know. So for me, I don't think that would be helpful. I think that would be really frustrating actually. So, I don't think I would find that helpful."*

And P8 explains:

> *"I think it could be useful but I think it could be a double-edged sword that, because you could set your expectations to that, and if it's not very high then your expectations would be lowered. However, it would be helpful in so much that maybe you could identify issues or problems before it started. I think it would definitely be a double-edged sword. And I think that a lot of staff around me, if they thought that somebody wasn't going to be engaging well, I don't think that they would put that much in."*

# APPENDIX 3: Interview Guide

1. **Demographics**
   - How did you become a SilverCloud supporter?
   - When did you start using this program (how many months/ years ago)?
   - How many clients are currently assigned to you to support through this program?
   - Considering your experience as a supporter of this program so far: would you describe yourself more as a novice, novice/ intermediate, intermediate/ expert or expert supporter?
   - For the reporting of this research, can I also ask you as what gender you identify as?

2. **Information Review + Response Process**
   - What do you like most about using the online CBT program to support clients?
   - Can you describe to me how you are supporting your clients? What do you do?
   - How do you get a sense of the person? How do you frame this person in your mind? How do you know? How do you connect to the person?
   - From the information currently available to you about a client, how easy or difficult is it for you to understand your clients' current situation?

3. **Personalising Messages**
   - When writing your feedback message, do you try and connect with the person on a personal level? How do you do that?
   - How do you choose what feedback provide to a particular client?
     - What information do you consider in your choice? Of this information, what would you say is most important? Why?
     - Once you selected a communication template, how much – if at all – do you adapt it before sending it out? How important is it for you to personalise your response message?
     - Going through your records, what would you consider to be a good example of a 'personalised message'? What about this message is more personal?
     - Do you think clients get a sense of you being a real person who is responding?
     - How well, do you think, you can express a sense of care or connection to the person? Is this something that's at all important to you?
     - What would you consider as most important in the information you currently have about a client in order for you to create a very personalised message?

4. **Additional Information Needs**
   - How confident are you that your response message was appropriate? Has there ever been a time, where you feel uncertain about what your response should be? Can you give an example?
   - Is there any information that would help you to support your clients better that is currently not (easily) available to you?
   - Do you have a sense as to why a client may not engage much with the platform? How much of such details do clients share in dialogue with you, if at all?